\documentclass[ALICE,manyauthors]{cernphprep}
\usepackage[comma,square,numbers,sort&compress]{natbib}
\usepackage{hyperref}

\usepackage{lineno}
\usepackage{xspace}
\usepackage{multirow}
\usepackage{graphicx}
\usepackage[loose]{units}
\usepackage{upgreek}
\usepackage{amsmath,amssymb,amsbsy,mathrsfs} 
\usepackage[T1]{fontenc}
\usepackage{orcidlink}

\begin{document}
%

\newcommand{\pp}           {pp\xspace}
\newcommand{\ppbar}        {\mbox{$\mathrm {p\overline{p}}$}\xspace}
\newcommand{\XeXe}         {\mbox{Xe--Xe}\xspace}
\newcommand{\PbPb}         {\mbox{Pb--Pb}\xspace}
\newcommand{\pA}           {\mbox{pA}\xspace}
\newcommand{\pPb}          {\mbox{p--Pb}\xspace}
\newcommand{\AuAu}         {\mbox{Au--Au}\xspace}
\newcommand{\dAu}          {\mbox{d--Au}\xspace}

\newcommand{\s}            {\ensuremath{\sqrt{s}}\xspace}
\newcommand{\snn}          {\ensuremath{\sqrt{s_{\mathrm{NN}}}}\xspace}
\newcommand{\pt}           {\ensuremath{p_{\rm T}}\xspace}
\newcommand{\xt}           {\ensuremath{x_{\rm T}}\xspace}
\newcommand{\meanpt}       {$\langle p_{\mathrm{T}}\rangle$\xspace}
\newcommand{\ycms}         {\ensuremath{y_{\rm CMS}}\xspace}
\newcommand{\ylab}         {\ensuremath{y_{\rm lab}}\xspace}
\newcommand{\etarange}[1]  {\mbox{$\left | \eta \right |~<~#1$}}
\newcommand{\yrange}[1]    {\mbox{$\left | y \right |~<~#1$}}
\newcommand{\dndy}         {\ensuremath{\mathrm{d}N_\mathrm{ch}/\mathrm{d}y}\xspace}
\newcommand{\dndeta}       {\ensuremath{\mathrm{d}N_\mathrm{ch}/\mathrm{d}\eta}\xspace}
\newcommand{\avdndeta}     {\ensuremath{\langle\dndeta\rangle}\xspace}
\newcommand{\dNdy}         {\ensuremath{\mathrm{d}N_\mathrm{ch}/\mathrm{d}y}\xspace}
\newcommand{\Npart}        {\ensuremath{N_\mathrm{part}}\xspace}
\newcommand{\Ncoll}        {\ensuremath{N_\mathrm{coll}}\xspace}
\newcommand{\dEdx}         {\ensuremath{\textrm{d}E/\textrm{d}x}\xspace}
\newcommand{\RpPb}         {\ensuremath{R_{\rm pPb}}\xspace}

\newcommand{\nineH}        {$\sqrt{s}~=~0.9$~Te\kern-.1emV\xspace}
\newcommand{\seven}        {$\sqrt{s}~=~7$~Te\kern-.1emV\xspace}
\newcommand{\twoH}         {$\sqrt{s}~=~0.2$~Te\kern-.1emV\xspace}
\newcommand{\twosevensix}  {$\sqrt{s}~=~2.76$~Te\kern-.1emV\xspace}
\newcommand{\five}         {$\sqrt{s}~=~5.02$~Te\kern-.1emV\xspace}
\newcommand{\twosevensixnn}{$\sqrt{s_{\mathrm{NN}}}~=~2.76$~Te\kern-.1emV\xspace}
\newcommand{\fivenn}       {$\sqrt{s_{\mathrm{NN}}}~=~5.02$~Te\kern-.1emV\xspace}
\newcommand{\LT}           {L{\'e}vy-Tsallis\xspace}
\newcommand{\GeVc}         {Ge\kern-.1emV/$c$\xspace}
\newcommand{\MeVc}         {Me\kern-.1emV/$c$\xspace}
\newcommand{\TeV}          {Te\kern-.1emV\xspace}
\newcommand{\GeV}          {Ge\kern-.1emV\xspace}
\newcommand{\MeV}          {Me\kern-.1emV\xspace}
\newcommand{\GeVmass}      {Ge\kern-.2emV/$c^2$\xspace}
\newcommand{\MeVmass}      {Me\kern-.2emV/$c^2$\xspace}
\newcommand{\lumi}         {\ensuremath{\mathcal{L}}\xspace}

\newcommand{\ITS}          {\rm{ITS}\xspace}
\newcommand{\TOF}          {\rm{TOF}\xspace}
\newcommand{\ZDC}          {\rm{ZDC}\xspace}
\newcommand{\ZDCs}         {\rm{ZDCs}\xspace}
\newcommand{\ZNA}          {\rm{ZNA}\xspace}
\newcommand{\ZNC}          {\rm{ZNC}\xspace}
\newcommand{\SPD}          {\rm{SPD}\xspace}
\newcommand{\SDD}          {\rm{SDD}\xspace}
\newcommand{\SSD}          {\rm{SSD}\xspace}
\newcommand{\TPC}          {\rm{TPC}\xspace}
\newcommand{\TRD}          {\rm{TRD}\xspace}
\newcommand{\VZERO}        {\rm{V0}\xspace}
\newcommand{\VZEROA}       {\rm{V0A}\xspace}
\newcommand{\VZEROC}       {\rm{V0C}\xspace}
\newcommand{\Vdecay} 	   {\ensuremath{V^{0}}\xspace}
\newcommand{\EMCal}        {\rm{EMCal}\xspace}
\newcommand{\DCal}         {\rm{DCal}\xspace}

\newcommand{\ee}           {\ensuremath{e^{+}e^{-}}} 
\newcommand{\pip}          {\ensuremath{\pi^{+}}\xspace}
\newcommand{\pim}          {\ensuremath{\pi^{-}}\xspace}
\newcommand{\kap}          {\ensuremath{\rm{K}^{+}}\xspace}
\newcommand{\kam}          {\ensuremath{\rm{K}^{-}}\xspace}
\newcommand{\pbar}         {\ensuremath{\rm\overline{p}}\xspace}
\newcommand{\kzero}        {\ensuremath{{\rm K}^{0}_{\rm{S}}}\xspace}
\newcommand{\lmb}          {\ensuremath{\Lambda}\xspace}
\newcommand{\almb}         {\ensuremath{\overline{\Lambda}}\xspace}
\newcommand{\Om}           {\ensuremath{\Omega^-}\xspace}
\newcommand{\Mo}           {\ensuremath{\overline{\Omega}^+}\xspace}
\newcommand{\X}            {\ensuremath{\Xi^-}\xspace}
\newcommand{\Ix}           {\ensuremath{\overline{\Xi}^+}\xspace}
\newcommand{\Xis}          {\ensuremath{\Xi^{\pm}}\xspace}
\newcommand{\Oms}          {\ensuremath{\Omega^{\pm}}\xspace}
\newcommand{\degree}       {\ensuremath{^{\rm o}}\xspace}


\newcommand{\piz}{$\pi^{0}$}

\newcommand{\ptg}  {\ensuremath{p_{\rm T}^{\gamma}}\xspace}
\newcommand{\xtg}   {\ensuremath{x_{\rm T}^{\gamma}}\xspace}
\newcommand{\xe}       {$x_{E}$}
\newcommand{\sigmalong}{\ensuremath{\sigma_{\rm long}^{2}}\xspace}

\newcommand{\evt} {$N_{\rm evt}$}
\newcommand{\lint}{$\mathcal{L}_{\rm int}$}

\newcommand{\alphaf}{$\alpha_{\rm MC}$}
\newcommand{\ptIso} {\ensuremath{p_{\rm T}^{\rm iso,~ch,~UE}}}
\newcommand{\ptT}   {\ensuremath{p_{\rm T}^{\rm trig}}}
\newcommand{\ptA}   {\ensuremath{p_{\rm T}^{\rm assoc}}}

\newcommand {\mom}  {\mbox{\rm  GeV$\kern-0.15em /\kern-0.12em c$}}
\newcommand {\gmom} {\mbox{\rm  GeV$\kern-0.15em /\kern-0.12em c$}}
\newcommand {\mass} {\mbox{\rm  GeV$\kern-0.15em /\kern-0.12em c^2$}}
\newcommand{\slfrac}[2]{\left.#1\right/#2}

\begin{titlepage}
\PHyear{2024}       
\PHnumber{171}      
\PHdate{21 June}  

\title{Measurement of the inclusive isolated-photon production cross section in \pp collisions at $\mathbf{\sqrt{\textit{s}}=13}$~TeV}

\ShortTitle{Isolated-photon production at \s~$=$~13~TeV in ALICE }   

\Collaboration{ALICE Collaboration\thanks{See Appendix~\ref{app:collab} for the list of collaboration members}}
\ShortAuthor{ALICE Collaboration} 

\begin{abstract}

The production cross section of inclusive isolated photons has been measured by the ALICE experiment at the CERN LHC in pp collisions at centre-of-momentum energy of $\s=13$~TeV collected during the LHC Run 2 data-taking period. 
The measurement is performed by combining the measurements of the electromagnetic calorimeter EMCal and the central tracking detectors ITS and TPC, covering a pseudorapidity range of $|\eta^{\gamma}|<0.67$  and a transverse momentum range of $7<\ptg<200~$\GeVc.  
The result extends to lower \ptg and $\xtg = 2\ptg/\s$ ranges,
the lowest \xtg\ of any isolated photon measurements to date,
extending significantly those measured by the ATLAS and CMS experiments towards lower \ptg\ 
at the same collision energy with a small overlap between the measurements.  
The measurement is compared with next-to-leading order perturbative QCD calculations and the results from the ATLAS and CMS experiments as well as with measurements at other collision energies. 
The measurement and theory prediction are in agreement with each other within the experimental and theoretical uncertainties.

\end{abstract}
\end{titlepage}

\setcounter{page}{2} 


\section{Introduction}

In high-energy hadronic collisions, direct photons are considered to be the most sensitive probe of the initial state. Direct photons are referred to as those photons which are emitted from the elementary processes, unlike decay photons produced from hadronic decays. A subset of direct photons is further classified as prompt photons which originate directly from the hard scattering of initial-state partons in hadronic collisions.
Direct prompt photons provide a probe to test perturbative Quantum Chromodynamics (pQCD) predictions and also constrain the parton distribution functions (PDF), in particular the gluon PDF~\cite{Arleo:2011gc,DENTERRIA2012311,Arleo:PhotonLHCYellow,Ichou:2010wc,Perez:2012um}.  
At leading order (LO) in pQCD, direct prompt photons are described via $2 \rightarrow 2$ processes: 
(i) quark--gluon Compton scattering, $\rm{q g} \rightarrow \rm{q} \gamma$,  (ii) quark--antiquark annihilation, $\rm{q \overline{q}} \rightarrow g \gamma$, and, with a much smaller contribution, $\rm{q \overline{q}} \rightarrow \gamma \gamma$. In addition, prompt photons are produced by next-to-leading order processes (NLO), like parton fragmentation or bremsstrahlung.
The collinear part of such processes has been shown to contribute effectively also at LO~\cite{Aurenche:1993}. A clean separation of the different prompt photon sources experimentally is difficult and with pQCD not possible, only the sum can be calculated, but one can suppress the contribution of fragmentation and bremsstrahlung, which are accompanied by other parton fragments, 
via the selection of ``isolated photons''. 
The sum of the transverse energies or transverse momenta (\pt) of the produced particles in a cone around the photon direction is required to be smaller than a given threshold value. The advantage of this selection is that it can be done both in the experimental measurement and theoretical calculations. 
Hence, isolation criteria should be applied to suppress fragmentation and bremsstrahlung photons while only marginally affecting direct prompt photon signal selection~\cite{Ichou:2010wc}. An isolation requirement is also important to reduce the background of decay photons, since hadrons at reasonably high \pt, which can decay to photons, are generally produced in jet fragmentation and are accompanied by additional jet fragments.

The high centre-of-momentum energy ($\sqrt s$) at the LHC allows one to access very small values of the longitudinal momentum fraction $x$ of the initial-state parton and \xt $=2p_{\rm T}/\sqrt{s}$, with $\xt\approx x$ at midrapidity, and thus constrain the PDFs~\cite{DENTERRIA2012311,ALICE:2019rtd}. 
In particular, the dominant contribution to the prompt photon production at the LHC is the quark--gluon Compton diagram~\cite{Ichou:2010wc}, which is directly sensitive to the gluon density, and has the largest uncertainty among the PDFs. 
Therefore, these $2 \rightarrow 2$ processes, in conjunction with the isolation selection criteria, probe the low-$x$ gluon content of one of the incoming protons. 

Measurements of direct photons and of isolated photons have been performed at the SPS~\cite{Appel:1986ix}, Tevatron~\cite{Abazov:2005wc,Aaltonen:2009ty}, and RHIC~\cite{PhysRevLett.98.012002} colliders and also earlier 
\cite{Ferbel:1984ef}. Different measurements were made at the LHC by the ATLAS and CMS Collaborations in pp collisions at various energies and they can be found in Refs.~\cite{Khachatryan:2010fm,CMS2011,Chatrchyan:2012vq,Sirunyan2019, Aad:2010sp,Aad:2011tw,Aad:2013zba,Aad:2016xcr,Aad:2017,Aad:2019,Aad:2023}. ALICE measured the isolated-photon yield in pp collisions at $\s =$~7~TeV~\cite{ALICE:2019rtd}
but also the direct-photon yield 
via the measurement of the excess above unity in the ratio of the inclusive-photon yield over the decay-photon yield 
in pp collisions at $\s =$~2.76 and 8~TeV~\cite{ALICE:2018mjj} and in Pb--Pb collisions at centre-of-momentum energy per nucleon pair $\sqrt{s_{\rm NN}} =$~2.76~TeV~\cite{ALICE:2015xmh}. ALICE also measured isolated photon--hadron correlations in pp and p--Pb collisions at $\sqrt{s_{\rm NN}} = 5.02$~TeV in~\cite{ALICE:2020atx}.

This paper presents the measurement of the isolated-photon cross section in pp 
collisions at $\s=$~13~TeV using a data sample collected with ALICE in years 2016, 2017, and 2018, with integrated luminosity 
$\mathscr{L}_{\rm int}=10.79$~pb$^{-1}$.
The measurements are performed in the photon transverse momentum range $7<~\ptg~<200$~\GeVc and correspondingly $1.1 \times 10^{-3}<\xt^{\gamma}<30.8 \times 10^{-3}$. 
This measurement follows closely the analysis strategy presented in the previous ALICE measurement in pp collisions at $\s=7$~TeV~\cite{ALICE:2019rtd} with two main differences, a larger calorimeter acceptance and the use of only charged particles in the isolation cone as discussed in the next sections.
This measurement covers a larger photon \pt\ range than the previous measurement, $10<\pt^{\gamma}<60$~\GeVc and $2.9 \times 10^{-3}<\xt^{\gamma}<~17.1 \times 10^{-3}$, due to the 20 times larger luminosity but also to the larger photon acceptance considered. It is worth noting that the lower-\ptg\ reach of the latest measurements of ATLAS~\cite{Aad:2019,Aad:2023} and CMS~\cite{Sirunyan2019} at $\s=13$~TeV is significantly higher than the measurement presented here, $125<\ptg<1000$~\GeVc and $190<\ptg<2500$~\GeVc, respectively, and thus, 
the lowest \xtg achieved by ATLAS and CMS
are $19.2 \times 10^{-3}$ and $29.2 \times 10^{-3}$, respectively. 
Thus, the ALICE result on isolated photons complements other LHC measurements by extending the kinematic range towards low \ptg and \xtg.

The paper is divided into the following sections: Section~\ref{sec:detector} presents the detector setup and data sample used for analysis; Section~\ref{sec:analysis} describes the analysis procedure; The systematic uncertainties are presented in Sect.~\ref{sec:sys_unc}; The final results and conclusions are presented in Sect.~\ref{sec:results} and~\ref{sec:conclusion}, respectively.


\section{Detector description and data selection\label{sec:detector}}

The ALICE experiment and its performance during the LHC Runs 1 and 2 
are described in Refs.~\cite{Aamodt:2008zz,Abelev:2014ffa}.
Photon reconstruction is performed using the Electromagnetic Calorimeter (EMCal)~\cite{ALICE:2022qhn} while charged particles used in the photon isolation selection criteria are reconstructed with the combination of the Inner Tracking System (ITS)~\cite{Aamodt:2010aa} and the Time Projection Chamber 
(TPC)~\cite{Alme:2010ke}, which are part of the ALICE central tracking detectors. 

The ITS is composed of six cylindrical layers of silicon detectors with full azimuthal acceptance and surrounds the interaction point. 
The two innermost layers consist of the Silicon Pixel Detector (SPD), whose fine granularity provides high-spatial precision for tracking close to the primary vertex. Those layers are positioned at radial distances of \unit[3.9]{cm} and \unit[7.6]{cm} from the beam line.
They are surrounded by the two layers of the Silicon Drift Detector (SDD) at \unit[15.0]{cm} and \unit[23.9]{cm}, and by those of the Silicon 
Strip Detector (SSD) at \unit[38.0]{cm} and \unit[43.0]{cm}. 
While the two SPD layers cover a pseudorapidity of $|\eta| <$ 2 and $|\eta| <$ 1.4, respectively, the SDD and the SSD subtend $|\eta| <$ 0.9 and $|\eta| <$ 1.0, respectively.
The TPC is a large ($\approx$ \unit[85]{m$^3$}) cylindrical drift detector filled with a gas mixture. 
It covers $|\eta| <$ 0.9 over the full
azimuth angle, with a maximum of 159 reconstructed space points along the track path.

The EMCal is a lead-scintillator sampling electromagnetic calorimeter used to measure photons, electrons, and the neutral part of jets via the electromagnetic showers that the different particles produce
in cells of the calorimeter. 
The scintillation light is collected by optical fibres coupled to Avalanche Photo Diodes 
(APD) that amplify the signal.
The energy resolution is $\sigma_E/E = A\oplus B/\sqrt{E}\oplus C/E$
with  $A = (1.4 \pm 0.1) \%$, $B = (9.5 \pm 0.2) \%$, $C = (2.9 \pm  0.9) \%$, 
and energy $E$ in units of GeV.
The EMCal was installed at a radial distance of \unit[4.28]{m} from the ALICE interaction point.
During the period in which the analysed dataset was collected, the EMCal consisted of twenty supermodules (SM). The supermodules have different sizes and are subdivided into $24 \times 48$ (along $\varphi \times \eta$) cells for ``full SM", $24 \times 32$ cells for ``2/3 SM", and  $8 \times 48$ cells for ``1/3 SM".
Ten ``full SM" and two ``1/3 SM" are installed 
with a total aperture of $|\eta|<0.7$ in pseudorapidity and $80^\circ < \varphi < 187^\circ$ in azimuthal angle. Six ``2/3 SM" are located facing the ``full SM" at $260^\circ < \varphi < 320^\circ$ with $0.22<|\eta|<0.7$, with a gap at $|\eta|<0.22$ covered by the PHOS detector~\cite{Acharya:2019rum} at $|\eta|<0.13$ (not used in this analysis) plus mechanical and electronics services. The remaining two ``1/3 SM" are located at $320^\circ < \varphi < 327^\circ$ with $|\eta|<0.7$. The eight SMs at the highest $\varphi$ including these last ``1/3 SM" and the ``2/3 SM" have previously been referred to as DCal.
Each cell has a transverse size of $6 \times 6~\mbox{cm}^2$ which corresponds to $\Delta \varphi \times \Delta \eta $ = 0.0143$\times$0.0143 rad, approximately twice the Moli\`ere radius. Thus, most of the energy of a single photon is deposited in a single cell plus the adjacent ones. 

The data were taken with a minimum bias interaction trigger (MB) and EMCal Level-1 photon-dedicated triggers (L1-$\gamma$).
The MB trigger was based on the response of the V0 detector, consisting of two arrays of 32 plastic scintillators, located at $2.8 < \eta < 5.1$ (V0A) and  $-3.7 < \eta < -1.7$ (V0C)~\cite{ALICE:2013axi}.
Two EMCal L1-$\gamma$ triggers were used, consisting in energy depositions larger than 4~GeV (L1-$\gamma$-low) or 9 GeV (L1-$\gamma$-high) in 4$\times$4 adjacent cells, 
in addition to the MB  trigger condition and an EMCal Level-0 (L0) trigger at 2.5~GeV, see detailed description of the EMCal triggers in Refs.~\cite{Bourrion_2013,ALICE:2022qhn}. 
All events with more than one reconstructed primary vertex were rejected in the analyses to exclude pileup events within the same bunch crossing and out-of-bunch pileup was removed with cuts on the V0 timing~\cite{Abelev:2014ffa}. Finally, only events with a primary vertex position along the beam direction within $\pm 10$~cm from the centre of the apparatus were considered in this analysis to grant a uniform pseudorapidity acceptance.
Table~\ref{tab:NevtRFLumi} lists the number of selected events per each of the triggers considered.

\section{Isolated-photon reconstruction and corrections}
\label{sec:analysis}
 
The analysis procedure followed to obtain the signal of isolated photons consists of the following steps: 
(a) reconstruction of clusters of cells from incoming particles in the calorimeter; 
(b) photon identification via track--cluster matching selection criteria and the study of the cell energy spread (shower shape) produced by the particles; and 
(c) selection of isolated photon clusters.

The detector response is modelled by Monte Carlo (MC) simulations reproducing the detector conditions of the data-taking period.
The corrections discussed in the next subsections are obtained using PYTHIA~8 (version 8.210~\cite{Sjostrand:2014zea} using the Monash 2013 tune~\cite{Skands_2014}) as particle generator. Two kinds of PYTHIA~8 processes are generated in intervals of the transverse momentum of the hard scattering, two jets (jet--jet) or a prompt photon and a jet 
($\gamma$--jet, mainly Compton and annihilation processes) as final state. 
The transport of the generated particles in the detector material is done using GEANT3~\cite{Geant3}. 
In the case of $\gamma$--jet event generation, the event 
is accepted when the prompt photon enters the EMCal acceptance. 
In the case of jet--jet event generation, the event is accepted when at least one jet 
produces a high-\pt\ photon originating from a hadron decay in the EMCal acceptance. 
This is done to enhance the number of such photons, which are the main 
background in this analysis. Besides, three sub-samples with different trigger thresholds ($\pt >3.5$ or 7~\GeVc or 28~\GeVc) have been used in the jet--jet event generation. 

In this section, the different steps to reconstruct, identify, and select isolated photons are explained. A more detailed description and discussion of the different items presented here can be found in Ref.~\cite{ALICE:2022qhn}. 

\subsection{Cluster reconstruction and selection \label{sec:clust}}
Particles deposit their energy in several calorimeter cells, forming a cluster. Clusters are obtained by grouping all cells with common sides 
whose energy is above 100\,MeV, starting from a seed cell with at least 500\,MeV. Furthermore, a cluster 
must contain at least two cells to ensure a minimum cluster size and to remove 
single-cell electronic noise fluctuations. Also, the highest-energy cell must be at a distance $d_{\rm mask}$ of at least two cells away from a known misbehaving cell to avoid possible electronic influence between channels, or from a dead cell to avoid holes in the cell energy spread of the clusters. 
To limit energy leakage at the SM borders, 
a distance of at least one cell of the highest-energy cell in the cluster to the SM border is required, except at the SM border at $\eta=0$ where two SMs are adjacent and clusters can be shared between SMs. 
These requirements lead to an acceptance depending on the configuration of SMs as presented in Table~\ref{tab:acceptance}.
Note that the acceptance in this measurement is 85\% larger than that in the $\s=$~7~TeV~\cite{ALICE:2019rtd} measurement that was limited to $|\eta|<0.27$ and $103^{\circ}<\varphi<157^{\circ}$ due to considerations on the isolation condition and geometry discussed later.

\begin{table}[htbp]
\begin{center}
\caption{\label{tab:acceptance} Measurement acceptance after cluster selection criteria depending on the calorimeter supermodule geometry. }
\setlength{\tabcolsep}{7mm}{
\begin{tabular}{ c c c}
SM type    & $\eta$ & $\varphi$ \\
\hline
Full SM    &  $|\eta|<0.67$      &  $81.2^{\circ}<\varphi<180^{\circ}$   \\        
2/3 SM     &  $0.25<|\eta|<0.67$ & $261.2^{\circ}<\varphi<318.8^{\circ}$ \\
1/3 SM     &  $|\eta|<0.67$      & $181.2^{\circ}<\varphi<185.8^{\circ}$   \\
1/3 SM     &  $|\eta|<0.67$      & $321.2^{\circ}<\varphi<325.8^{\circ}$ \\
\hline

\end{tabular}}

\end{center}
\end{table}

During the Run 2 data-taking periods with pp collisions at \s~=~13~TeV, the LHC delivered events in bunches separated by 25\,ns with an average number of collisions per bunch crossing changing over the years, being about $\mu = 0.005-0.02$. 
The EMCal time resolution is between 1 and 2 ns below cluster energy $E=80$~GeV, reaching close to 3 ns above 100 GeV~\cite{ALICE:2022qhn}. 
To ensure the selection of clusters from the main bunch crossing, 
the measured time of the highest-energy cell in the clusters relative to the main bunch crossing has to satisfy $\Delta t<20$~ns. 

An energy non-linearity correction derived from electron test beam data~\cite{ALICE:2022qhn} is applied to the reconstructed cluster energy. It amounts to about 6\% at $E=1$~GeV decreasing to approximately 2\% between 20 and 100~GeV and increasing to approximately 3\% at 200 GeV due to signal saturation. 
Since the simulation does not exactly reproduce the energy linearity, an extra correction factor is applied to match the $\pi^{0}$ mass obtained in simulations to the corresponding measurements in data. The value of this correction is 1.025 at $E=1$~GeV, decreasing to 0.975 above 7~GeV. After the corrections, the energy scale uncertainty of the EMCal is considered to be 0.5\%.

Nuclear interactions occurring in the APD, in particular those involving neutrons, induce an abnormal signal~\cite{ALICE:2022qhn,Bialas:2013wra}. 
Such a signal is most frequently observed as a single high-energy cell
with no or few surrounding low-energy cells, 
this latter case being mostly caused by cell cross talk or overlaps of the underlying event particles and this abnormal signal in the measurement.
The abnormal signals can be removed by comparing
the energies in adjacent cells to the cell with maximum energy
$E_{\rm max}$. To reject these signals, one requires that the
ratio $F_{+} \equiv 1-E_{+}/E_{\rm max}$, where $E_{+}$ is the
sum of the energy of the four surrounding cells that share a
common edge with the maximum cell, satisfies $F_{+} < 97$\%. 
However, cross talk is happening to cells in the same readout
card, called T-Card, which serves $2\times 8$ cells along
$\eta\times\varphi$. Cross talk of the abnormal signal can
cause surrounding cells with higher signals than otherwise
expected, such that, at very high cluster energies, the selection
using $F_{+}$ is not sufficient to reject those signals. 
Since clusters with high energy from physical signals must contain cells
from more than one T-Card~\cite{ALICE:2022qhn}, 
this additional condition is required for clusters with $E\geq 80$ GeV to reject the abnormal signals.

Contamination of the cluster sample by charged particles is suppressed by a charged particle veto (CPV). Tracks of charged particles are reconstructed in a hybrid approach using ITS and TPC, which reduces local inefficiencies potentially caused by non-functioning elements of the ITS.
Two distinct track classes are accepted in this method~\cite{Abelev:2014ffa}: (i) tracks containing at least three hits in the ITS, including at least one hit in the SPD, with 
momentum determined from a Kalman-filter fit to the hits attached to the track,
and (ii) tracks containing less than three hits in the ITS or no hit in the SPD, with the primary vertex included in the momentum determination. Class (ii) is 
used only when SPD modules along the particle trajectory are inactive.
Class (i) contributes 90\% and class (ii) 10\% of all accepted tracks, independently of \pt.  
Charged-particle tracks are selected to have a distance of closest approach 
to the primary vertex smaller than 2.4\,cm in the plane transverse 
to the beam, and smaller than 3.0\,cm in the beam direction.
EMCal clusters originating from charged particles are tagged by applying a selection on the separation of the position of the track extrapolated to the EMCal surface from the cluster position, which must fulfil
\begin{equation} 
\label{eq:cpv}
 \Delta \eta^\text{residual} < 0.010 +(\pt^\text{track} +4.07)^{-2.5}\ \mathrm{and}\ \Delta \varphi^\text{residual} < 0.015 +(\pt^\text{track} +3.65)^{-2} ~\mathrm{rad}
\end{equation}

where 
$\Delta \eta^\text{residual}=|\eta^\text{track}-\eta^\text{cluster}|$, $\Delta \varphi^\text{residual}=|\varphi^\text{track}-\varphi^\text{cluster}|$, 
and the track transverse momentum ($\pt^\text{track}$) is in \GeVc\ units.
This condition is applied if the ratio of cluster energy over track momentum is smaller than 1.7, used to reduce the amount of fake vetoes, if larger, no cluster--track association is considered~\cite{ALICE:2022qhn}.
The track-to-cluster matching efficiency amounts to about 92\% for primary charged hadrons and electrons at cluster energies of $E \simeq 1$\,GeV, and increases up to 96\% for clusters of 10\,GeV. 

From now on, clusters that are not matched to any charged-particle track are called ``neutral clusters''.

\subsection{Photon identification via cluster shower shape measurement}
\label{sec:photonident}

The neutral clusters can have a wider elongated shape if one or several additional 
particles deposit their energy nearby in the detector. 
The most frequent case is a two-particle cluster when the distance between particles is larger than two cells. In such cases, 
one can observe clusters with more than one local maximum in the energy distribution, where a local maximum is defined as a cell with a signal higher than 
the neighbouring cells. 

For an increasing number of local maxima ($N_{\rm LM}$),
the cluster will in general get wider. 
Prompt photons  generate clusters with $N_{\rm LM}=1$, except if they suffer conversion in the material in front of the EMCal. 
The two decay photons from 
high-\pt\ \piz\ and  $\eta$ mesons with energy 
above 6 and 24\,GeV, respectively, likely merge into a single cluster as observed in simulations.
Merged clusters from \piz\ mesons with energy below 15~GeV and $\eta$ mesons with energy below 60\,GeV most often have 
$N_{\rm LM} = 2$. With increasing energy, the two-photon opening angle decreases, leading to 
merged clusters with mainly $N_{\rm LM} = 1$ above 25\,GeV for \piz\ mesons and above 100\,GeV for $\eta$ mesons. 
Clusters with $N_{\rm LM}>2$ are rejected in this analysis, as these clusters are the major contribution to the background and clusters produced by more than two particles are not perfectly reproduced in Monte Carlo simulations. 
The contribution of clusters with $N_{\rm LM}=2$ is especially large in the case of wide showers, and is crucial for the estimate of the contamination of the prompt-photon sample, as explained in Sect.~\ref{sec:purity}. 

Merged and single photon clusters can be discriminated based on the shower shape using 
the width parameter \sigmalong, i.e. the square of the larger eigenvalue of the energy distribution in the $\eta-\varphi$ plane~\cite{ALICE:2022qhn},
that can be calculated as

\begin{equation}
\label{eq:ss1}
\sigma_{\rm long}^{2} = (\sigma_{\varphi\varphi}^{2}+\sigma_{\eta\eta}^{2})/2+\sqrt{(\sigma_{\varphi\varphi}^{2}-\sigma_{\eta\eta}^{2})^{2} / 4+\sigma_{\eta\varphi}^{4}},\\
\end{equation}

where ${\sigma^2_{xz}} = \big \langle xz \big \rangle - \big \langle x \big \rangle \big \langle z \big \rangle$ and $\big \langle x \big \rangle = (1/w_{\rm tot}) \sum w_i x_i$ 
are weighted over all cells associated with the cluster in the $\varphi$ or $\eta$ direction. The weights $w_i$ depend logarithmically on the ratio of the energy 
$E_i$ of the $i$-th cell 
to the cluster energy, as $w_i = \mathrm{max}(0,4.5+\ln(E_{i}/E_{\rm cluster}))$ and 
$w_{\rm tot} = \sum w_i$~\cite{Awes:1992}.

\begin{figure}[tb]
\begin{center}
    \begin{minipage}{0.51\textwidth}
    \begin{center}
    \subfigure[]{
        \label{fig:2DM02}
        \includegraphics[width=0.97\textwidth]{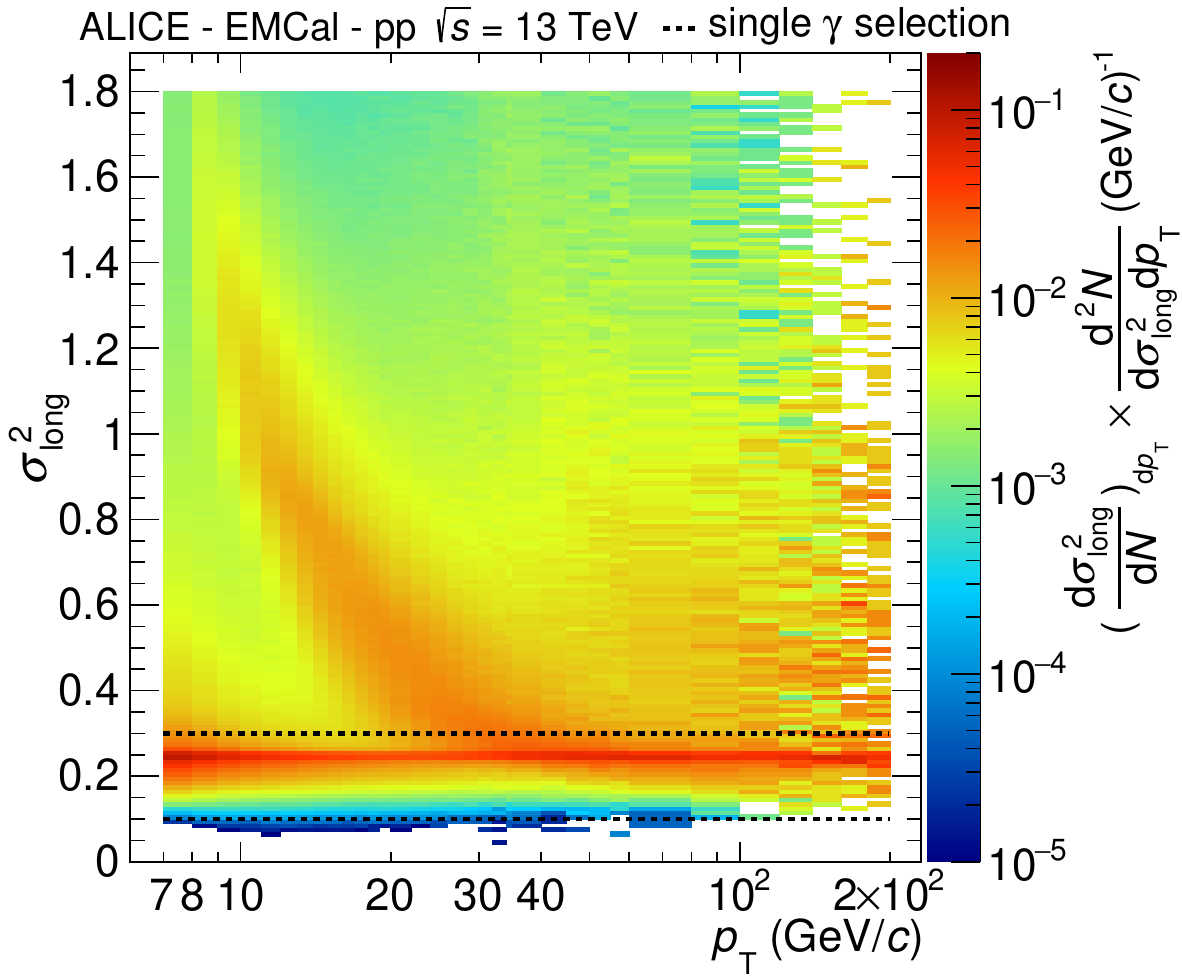}
    }
    \end{center}
    \end{minipage}
    \begin{minipage}{0.48\textwidth}
    \begin{center}
    \subfigure[]{
        \label{fig:1DM02}
        \includegraphics[width=1.\textwidth]{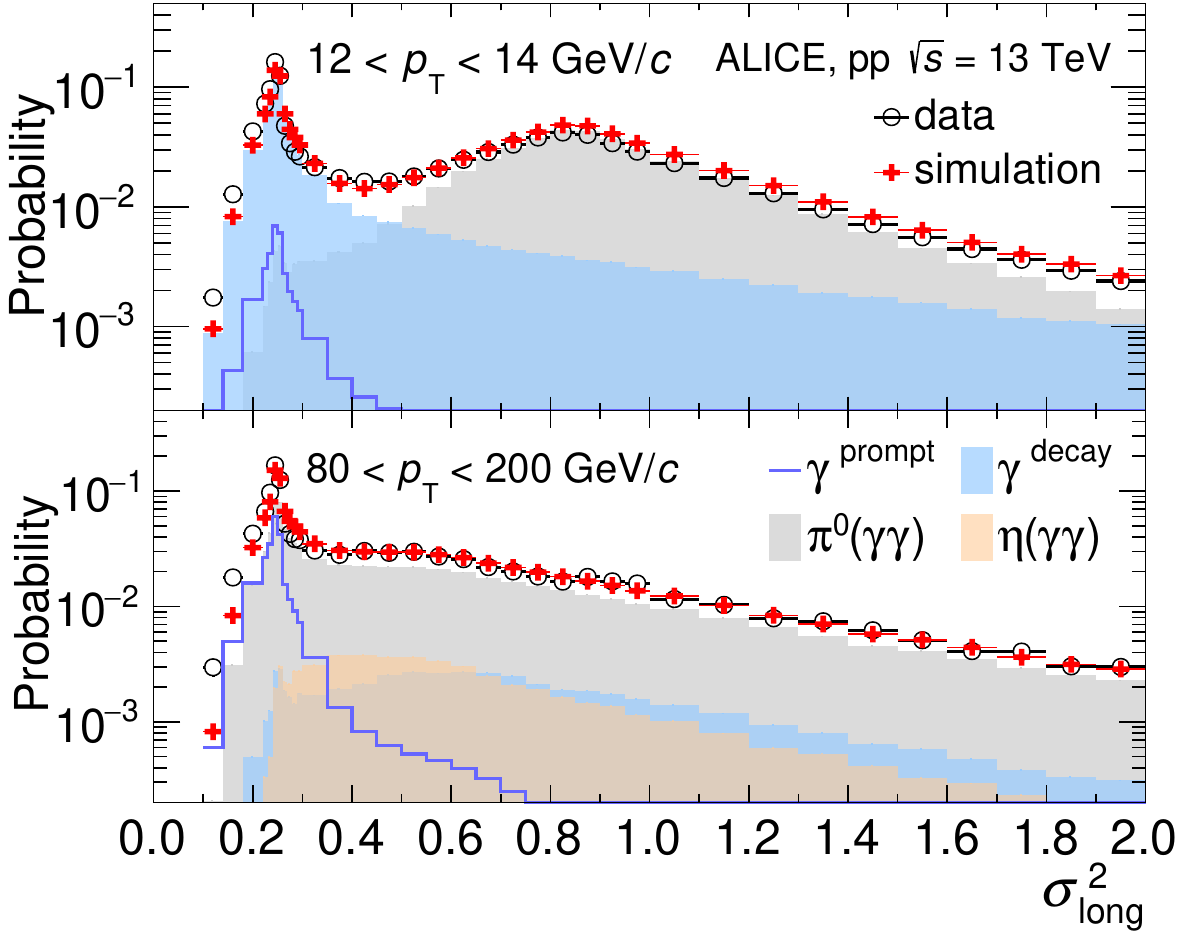}
    }
    \end{center}
    \end{minipage}
\end{center}
\caption{\small {(colour online) (a) Cluster shower shape distribution \sigmalong\ as a function of the cluster \pt\ in data. The dashed lines correspond to the upper and lower selection limit values for single photon candidate clusters used in the analysis. 
(b) Projected shower shape distribution in two \pt\ intervals in data (black circles) and simulation (red crosses, PYTHIA~8 jet--jet and $\gamma$--jet events). For the simulation, the \sigmalong\ distribution for clusters with different particle origins are also shown as shaded areas except for prompt photons from $\gamma$--jet PYTHIA~8 processes that are presented with a blue histogram line. 
}}
\label{fig:M02}
\end{figure}

The neutral-cluster \sigmalong\ distributions as a function of the cluster \pt\ are shown 
in Fig.~\ref{fig:2DM02} for data.
Most of the pure single photons are reconstructed as clusters with \sigmalong~$\approx 0.25$, other cases contribute to higher values, as seen in Fig.~\ref{fig:1DM02} where data and simulation distributions are compared.
In this analysis, “photon candidates” refer to clusters with a narrow shape defined by
$0.1 <$~\sigmalong~$< 0.3$. 
Above the higher limit in \sigmalong, defined by the dashed line in Fig.~\ref{fig:2DM02},  a clear \pt-dependent band is observed. This band is populated by two \piz-decay photons contributing to a single cluster as it can be deduced from  Fig.~\ref{fig:1DM02},  which shows different particle contributions to the shower shape distribution for two neutral-cluster \pt\ intervals. The dominant contributions to the narrow shower shape region are from single photons (at low \pt) or merged photons from neutral meson decays (at high \pt).
Another more faint band appears in Fig.~\ref{fig:2DM02} for $\pt>40$~\GeVc due to merged clusters from $\eta$ meson decays, as it can be seen from the simulations reported on the lower panel of Fig.~\ref{fig:1DM02}.  
The value of \sigmalong\ for merged photon clusters from meson decays decreases with increasing \pt, which leads to an almost full overlap of the \piz-decay photon band with the single photon shower band for about $\pt>40$~\GeVc. 
The lower limit at \sigmalong~$=0.1$ is set to clean the cluster sample from a few anomalous high-energy depositions that still pass the $F_{+}$ selection.

The cross talk between cells mentioned above modifies the shower shape distribution by widening the single-photon-cluster peak distribution, especially on the right tail of Fig.~\ref{fig:1DM02} for prompt photons, as discussed in detail in Ref.~\cite{ALICE:2022qhn}. 
This effect was modelled in simulation adding a small fraction of energy of the highest-energy cell into the surrounding cells in the same T-Card, and a good agreement between data and simulation was achieved, as seen in Fig.~\ref{fig:1DM02}.

\subsection{Isolated-photon selection}
\label{sec:isolation}

Direct prompt photons are mostly  isolated, i.e.\ have no hadronic activity in their vicinity except for the underlying event of the collision, in contrast to other photon sources like photons from parton fragmentation or decays of hadrons which have a high probability to
be accompanied by other fragments ~\cite{Ichou:2010wc}. 

An isolation criterion is applied to the photon candidate to suppress the contribution by fragmentation and decay photon production.
An equivalent isolation criterion is commonly included in theoretical calculations to account for the suppression of the fragmentation contribution to the total prompt photon cross section.

The isolation criterion is based on the so-called ``isolation momentum'' $p_{\rm T}^{\rm iso}$, i.e. the transverse momentum sum of all particles measured inside a cone of radius $R$ around the photon candidate, located at $\eta^{\gamma}$ and $\varphi^{\gamma}$. 
A particle of coordinates $\eta$ and $\varphi$ in angular space is inside the cone when 

\begin{equation}
\label{eq:rsize}
\sqrt{ (\eta - \eta^{\gamma})^2 + (\varphi-\varphi^{\gamma})^2} < R.
\end{equation}

The cone radius $R=0.4$ is chosen as it contains the dominant fraction of the jet energy~\cite{PhysRevD.71.112002}. 

In this analysis, the isolation momentum is the sum of the transverse momenta of all of the 
charged tracks (ch) that fall into the cone
\begin{equation}
\label{eq:etiso}
p_{\rm T}^{\rm iso,~ch,~UE}=\sum p_{\rm T}^{\rm track},
\end{equation}
where UE indicates that the tracks of the charged particles from the underlying event are not subtracted.
The candidate photon is declared isolated if $p_{\rm T}^{\rm iso,~ch,~UE}<$~1.5\,\GeVc. 
The contribution from UE tracks in the isolation cone was considered to be small, a contribution smaller than 1~\GeVc on average. Nevertheless, the effect they have in the selection is corrected, as discussed later. Charged particles used in the calculation of the isolation momentum are from the same track classes used for the track--cluster matching presented earlier.
Accepted tracks in the cone satisfy $|\eta^{\rm track}|< 0.9$ and $\pt^{\rm track}>0.15$~\GeVc. 

In the previous measurement in pp collisions at \s~=~7~TeV~\cite{ALICE:2019rtd}, 
the same $R=0.4$ radius value was used, but with a different definition of  the isolation momentum:
the transverse momenta of neutral clusters in the calorimeter, excluding the candidate photon, were also included.
The price to pay was to heavily reduce the acceptance of the analysis to $|\eta|<0.27$ to ensure that the cone was fully contained in the calorimeter acceptance. It was found that the use of tracks or tracks plus calorimeter clusters was equivalent once the isolation momentum threshold selection was adjusted from $p_{\rm T}^{\rm iso,~UE} < 2$~GeV/$c$, used in the previous analysis, to $p_{\rm T}^{\rm iso,~ch,~UE} < 1.5$~GeV/$c$. Thus, we decided to use only tracks, like in the measurement of isolated photon--hadron correlations in pp and p--Pb collisions  at $\sqrt{s_{\rm NN}} = 5.02$~TeV reported in Ref.~\cite{ALICE:2020atx}. 
However, in this measurement, when the cluster candidate for isolation is at $0.5<|\eta|<0.67$, a small fraction of the isolation cone is out of the tracking acceptance. 
To have the largest acceptance possible, and thus, the largest amount of photons, such candidate clusters are accepted in the analysis. For these cases, the measured isolation momentum with tracks is scaled up by the fraction of the cone area that is out of the acceptance, as in Ref.~\cite{ALICE:2020atx}.

\subsection{Purity of the isolated-photon sample\label{sec:purity}}

The isolated-photon candidate sample still has a non-negligible contribution from background clusters, mainly from neutral-meson decay photons. 
To estimate the background contamination, the same procedure as in Ref.~\cite{ALICE:2019rtd} is followed. Different classes of measured clusters 
were used:
(1) classes based on the shower shape \sigmalong, i.e.\ \textit{narrow} (photon-like) and \textit{wide} (most often elongated, i.e.\ non-circular), and 
(2) classes defined by the isolation momentum $p_{\rm T}^{\rm iso}$, i.e.\ \textit{isolated} (iso) and \textit{anti-isolated} ($\overline{\rm iso}$).  
The different classes are denoted 
by sub- and superscripts, e.g. isolated, narrow clusters are given as $X_{\rm n}^{\rm iso}$ and anti-isolated, wide clusters are given as $X_{\rm w}^{\overline{\rm iso}}$.
The wide clusters (mostly background) correspond to clusters with $ 0.4 <~$\sigmalong~$ <2.4$ and narrow clusters (containing most of the signal) are defined in Sect.~\ref{sec:photonident}. 
The isolation criterion corresponds to $p_{\rm T}^{\rm iso,~ch,~UE}<$~1.5\,\GeVc\ whereas the anti-isolation corresponds to $2.5<p_{\rm T}^{\rm iso,~ch,~UE}<$~10\,\GeVc. 
The yield of isolated-photon candidates in this nomenclature is $N_{\rm n}^{\rm iso}$. 
It consists of signal ($S$) and background ($B$) contributions: 
$N_{\rm n}^{\rm iso} = S_{\rm n}^{\rm iso} + B_{\rm n}^{\rm iso}$.
The contamination of the candidate sample is then $C = B_{\rm n}^{\rm iso}/N_{\rm n}^{\rm iso}$, or respectively, the purity is then $P \equiv 1 - C$.
Assuming that the ratios of the isolated over the anti-isolated  background for narrow clusters is the same as for wide clusters so that
\begin{equation}
\label{eq:ABCDproportionality}
\frac{B_{\rm n}^{\rm iso}/B_{\rm n}^{\overline{\rm iso}}}{B_{\rm w}^{\rm iso}/B_{\rm w}^{\overline{\rm iso}}}=1,
\end{equation}
and assuming that the proportion of signal in the control regions is negligible, the purity is derived in a data-driven approach (dd) as
\begin{equation}
\label{eq:ABCDpurity}
P_{\rm dd}=1-\frac {B_{\rm n}^{\overline{\rm iso}}/N_{\rm n}^{\rm iso}} {B_{\rm w}^{\overline{\rm iso}}/B_{\rm w}^{\rm iso}} = 1-\frac {N_{\rm n}^{\overline{\rm iso}}/N_{\rm n}^{\rm iso}} {N_{\rm 
w}^{\overline{\rm iso}}/N_{\rm w}^{\rm iso}}.
\end{equation}
Unfortunately, both assumptions are valid only approximately, especially Eq.~\eqref{eq:ABCDproportionality}. In PYTHIA~8 simulations with two jets in the final state that contribute only to the background in all of the four 
classes, an evaluation of Eq.~\eqref{eq:ABCDproportionality} gives values of the 
order of 1.1 at \ptg~=~10--40\,\GeVc, 
decreasing to about 0.7 for \ptg~$\approx 200 $\,\GeVc, 
thus the ratio is in general different from unity.
Since these deviations from unity are purely due to particle kinematics and detector response, the simulation can be used to estimate the bias via
\begin{equation}
\label{eq:ABCDproportionalityMC}
\bigg ( \frac{B_{\rm n}^{\rm iso}/B_{\rm n}^{\overline{\rm iso}}}{B_{\rm w}^{\rm iso}/B_{\rm w}^{\overline{\rm iso}}} \bigg)_{\rm data}= \bigg ( \frac{B_{\rm n}^{\rm iso}/B_{\rm n}^{\overline{\rm 
iso}}}{B_{\rm w}^{\rm iso}/B_{\rm w}^{\overline{\rm iso}}} \bigg)_{\rm MC}.
\end{equation}

This implies replacing  Eq.~\eqref{eq:ABCDproportionality} by the relation given by Eq.~\eqref{eq:ABCDproportionalityMC} leading to the expression of the MC-corrected purity 

\begin{equation}
\label{eq:ABCDpurityMC}
P = 1-\bigg(\frac {N_{\rm n}^{\overline{\rm iso}}/N_{\rm n}^{\rm iso}} {N_{\rm w}^{\overline{\rm iso}}/N_{\rm w}^{\rm iso}}\bigg)_{\rm data} \times \bigg(\frac {B_{\rm n}^{\rm iso}/N_{\rm 
n}^{\overline{\rm iso}}} {N_{\rm w}^{\rm iso}/N_{\rm w}^{\overline{\rm iso}}}\bigg)_{\rm MC}  \equiv 1-\bigg(\frac {N_{\rm n}^{\overline{\rm iso}}/N_{\rm n}^{\rm iso}} {N_{\rm w}^{\overline{\rm 
iso}}/N_{\rm w}^{\rm iso}}\bigg)_{\rm data} \times \alpha_{\rm{MC}},
\end{equation}

where MC contains both jet--jet and $\gamma$--jet events scaled to their respective cross sections.

The difference between the degree of the correlation between isolation momentum and shower shape distribution in data and simulation is another potential source of bias, as it influences the validity of Eq.~\eqref{eq:ABCDproportionalityMC}. 
To check this, the dependence of the double ratio
\begin{equation}
\label{eq:doubleratio}
\frac {\left( N^{{\rm iso}}/N^{\overline{\rm iso}} \right) ^{\mathrm{data}}}
{\left( N^{{\rm iso}}/N^{\overline{\rm iso}} \right) ^{\mathrm{MC}}} = 
f\left( \sigma_{\rm long}^{2} \right)
\end{equation}
on the shower shape width \sigmalong\ is studied in a region where the signal contribution is expected to be negligible. 
If the correlation between the two variables is correctly reproduced in the simulation, the double ratio is independent of \sigmalong, i.e.\ it would be the same for wide and narrow clusters. 
The double ratio was found to be above unity, indicating a larger isolation probability in data than in simulations. 
This is mainly due to an imperfect  calibration of charged particle tracks which leads to some discrepancy between data and simulations in the estimate of the isolation energy from charged particles. 
However, since the correction introduced in Eq.~\eqref{eq:ABCDpurityMC} relies on a narrow-over-wide ratio, the overall normalisation in the double ratio of Eq.~\eqref{eq:doubleratio} does not enter 
the correction.

The double ratio $f(\sigma_{\rm long}^{2})$ was found to have a small slope depending on \sigmalong\ which changes for the different \pt-intervals of the measurement.
A possible bias has been estimated via extrapolations by linear fits of the dependence on \sigmalong\ instead of the original assumption of a constant value. This consists of replacing the MC correction in Eq.~\eqref{eq:ABCDpurityMC} by a modified term
\begin{equation}
\label{eq:CorrectionMCExtra}
\alpha_{{\rm MC}} \longmapsto \alpha_{{\rm MC}} \times \left(\frac{p_0+\sigma^2_\mathrm{long,n} \times p_1}{p_0+\sigma^2_\mathrm{long,w} \times p_1}\right),
\end{equation}
where $\sigma^2_\mathrm{long,n}$ and $\sigma^2_\mathrm{long,w}$ are the median values of the neutral-cluster \sigmalong\ distribution in the narrow and wide ranges, respectively, and $p_0$ and $p_1$ are 
the parameters of the linear fit of the double ratio $f(\sigma_{\rm long}^{2})$.
These extrapolations have then been used in the estimate of the uncertainties of the purity. When referring to the systematic uncertainty in the following, this contribution is called isolation probability.
 
\begin{figure}[ht]
\begin{center}
\includegraphics[width=0.7\textwidth]{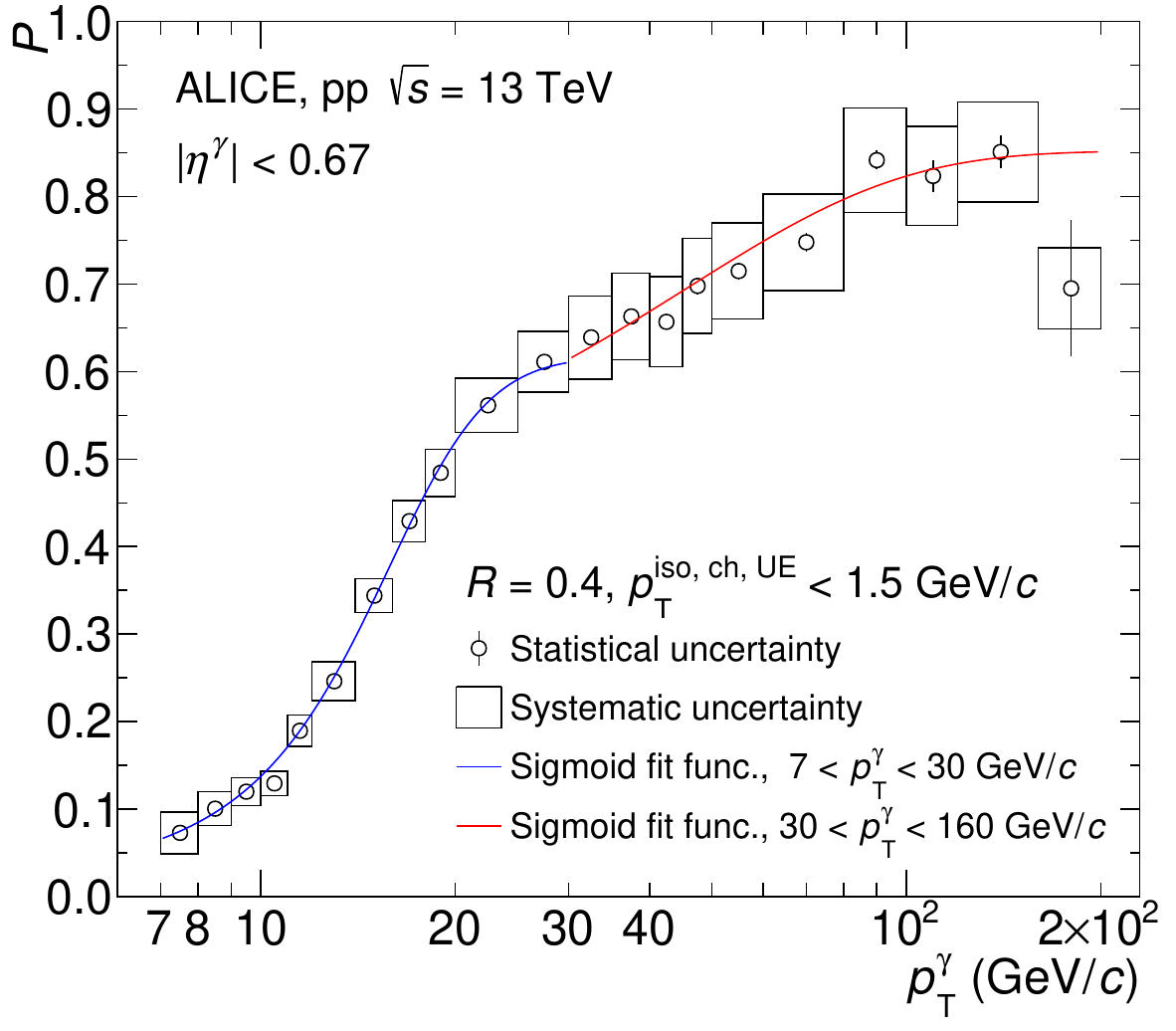}
\end{center}
\caption{\label{fig:purityIsoPhoton} (colour online) 
Purity of the isolated-photon sample as a function of \ptg\ calculated using Eq.~\eqref{eq:ABCDpurityMC}. The statistical and systematic uncertainties have been evaluated as discussed in Sect.~\ref{sec:sys_unc}. The red and blue lines are the results of a fit to the measured purity using a two-sigmoid-function described by Eq.~\eqref{eq:sigmoid} for two different transverse momentum intervals.
}
\end{figure}

Figure~\ref{fig:purityIsoPhoton} shows the purity calculated using Eq.~\eqref{eq:ABCDpurityMC}. 
The boxes indicate the systematic uncertainty whose estimation is explained in 
the next section that includes the variation of Eq.~\eqref{eq:CorrectionMCExtra}. 
There is a large contamination at $\ptg = 7$--10\,\GeVc\ of 95--90\% that decreases and partially saturates at 40--50\% for  $\ptg > 18$\,\GeVc. 
It decreases again above 40\,\GeVc 
and shows hints of a further saturation above 80\,\GeVc at 20\%.
The purity is comparable to the previous ALICE isolated-photon measurements in pp collisions at $\s=5.02$ and 7~TeV~\cite{ALICE:2019rtd,ALICE:2020atx}.

The \ptg\ dependence of the purity is caused by an interplay of physics and detector effects. 
Most of the contamination is due to \piz-decay photons.
On the one hand, the \pt\ spectra of prompt photons are harder 
than those of neutral pions, mainly because the latter undergo fragmentation, as also was found in pQCD calculations~\cite{Abeysekara:2010ze,Arleo:PhotonLHCYellow}.
For this reason, the $N_{\gamma_{\rm 2 \rightarrow 2}} / N_{\gamma(\pi^{0})}$  yield ratio rises with \ptg, and the photon purity increases as well. 
Also, the probability of tagging a photon as isolated varies with \ptg. 
At higher decay-photon \ptg, isolation is less probable for a fixed isolation momentum. 
On the other hand, the rejection of clusters from \piz\ and $\eta$ decays 
at high \pt becomes less effective due to the decreasing decay-photon opening angle when increasing the meson \pt. 
Below 18\,\GeVc, the contamination is dominated by single (i.e.\ unmerged) decay photons from {\piz} mesons, the remaining contributors being mainly photons from $\eta$ meson decays. 
Above 18\,\GeVc, a large fraction of the $\pi^{0}\rightarrow\gamma\gamma$ decays produces two photons with narrow opening angle and gives rise to merged clusters in the EMCal with a narrow shower shape that satisfies the condition for the single photon signal, as can be appreciated in Fig.~\ref{fig:2DM02}. 
In the PYTHIA 8 jet--jet simulations, at $\pt = 14$--16\,\GeVc approximately 6\% of reconstructed clusters in the narrow shower shape region come from \piz\ clusters, then at $\pt = 20$--25\,\GeVc, they are more than 30\%  and above 60\,\GeVc this contribution rises to about 80\%. 
The clusters produced by merged photons from $\eta$-meson decays contribute to the narrow shower shape region for $\pt>60$~\GeVc but they remain subdominant compared to merged \piz-decay clusters. Instead, in the range $40<\pt<60$~\GeVc, most of the merged $\eta$-decay clusters have wide shower shapes, which is in part the reason for the increase of purity in this \ptg\ region since the contribution of single photon clusters from $\eta$ decays to the narrow clusters decreases.
The combined effect of these mechanisms leads to the rise of the purity at low \pt,  saturation for $18<\ptg <  40$\,\GeVc, then rise above 40~\GeVc, and finally saturation above 80 \GeVc.

To reduce the  point-to-point statistical fluctuations in the purity used to correct the isolated-photon raw yield, the distribution is fitted by two sigmoid functions to reproduce the trend of the distribution with \ptg 
\begin{equation} \label{eq:sigmoid}
f_{i, ~\rm fit-sig}(\ptg) = \frac{a_i}{1+\exp(-b_i \times (\ptg - c_i))},
\end{equation}
where $i$ indicates the different fitting ranges. The first fit is done from $\ptg =7$ to 30~\GeVc\ with fit parameters obtained being $a_{1} = 0.617\pm 0.003$, $b_{1} = 0.292\pm 0.002$, and $c_{1} = 14.28\pm 0.05$. 
The second fit is done from $\ptg =20$ to 160~\GeVc\ with parameters 
$a_{2} = 0.852\pm 0.014$, $b_{2} = 0.034\pm 0.003$, and $c_{2} = 2.4\pm 1.0$.
The fit results are shown in the blue and red lines of Fig.~\ref{fig:purityIsoPhoton}, depending on the fit \ptg range, the first fit is used from $\ptg =7$ to 30 GeV/$c$ and the second from $\ptg =30$ to 200~\GeVc.

\subsection{Isolated-photon efficiency\label{sec:efficiency}}

The photon reconstruction, identification and isolation  efficiency has been computed using PYTHIA~8 simulations of $\gamma$--jet processes
in which, for each event, a prompt photon from $2\rightarrow 2$ Compton or annihilation processes (also two photons in the final state but negligible contribution) is emitted in the EMCal acceptance. Only 
the photons falling in the fiducial acceptance are considered in the efficiency calculation. 

\begin{figure}[ht]
\begin{center}
    \begin{minipage}{0.48\textwidth}
    \begin{center}
    \subfigure[ ]{
        \includegraphics[width=1.\textwidth]{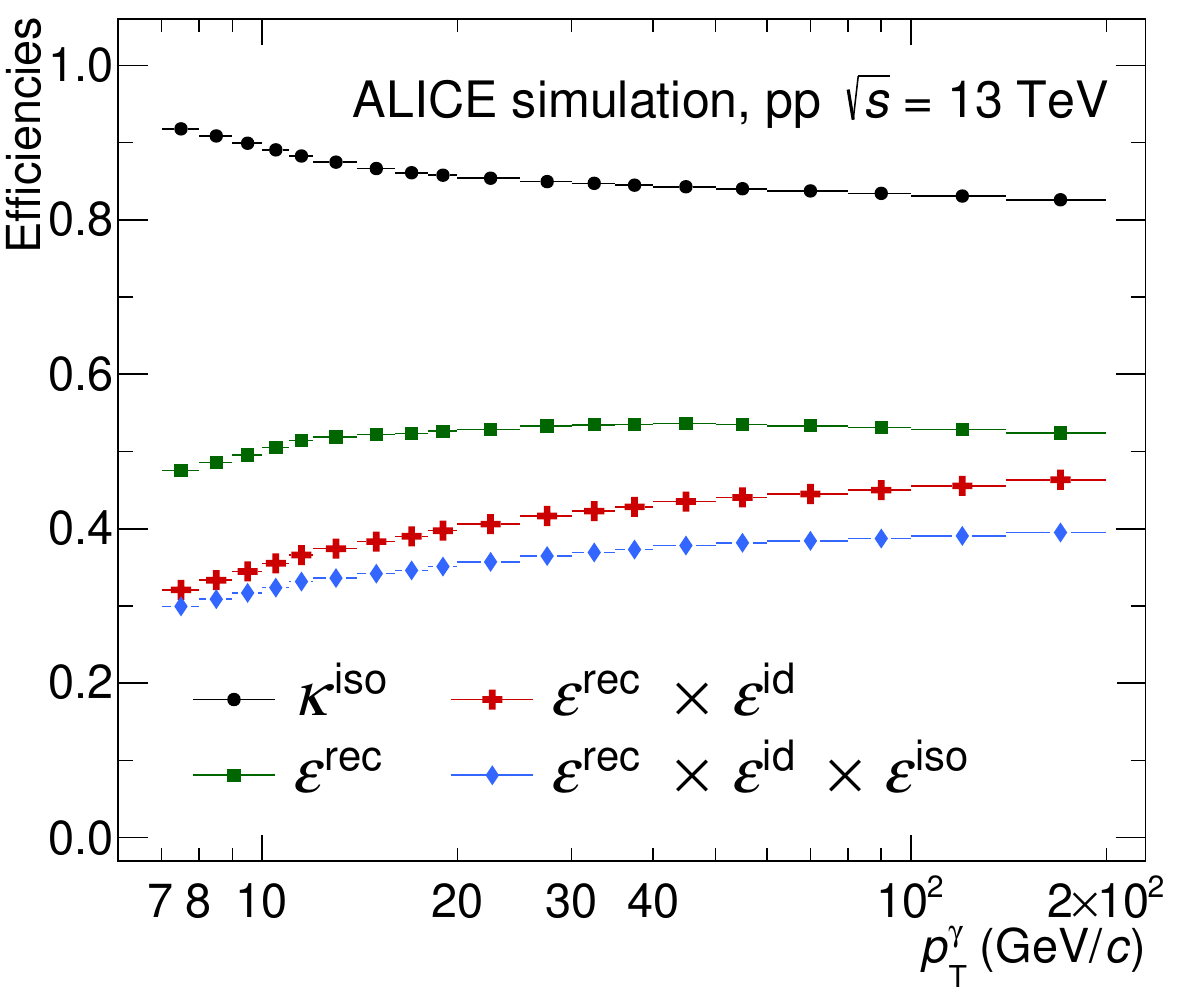}
        \label{fig:effiComponent}
    }
    \end{center}
    \end{minipage}
    \begin{minipage}{0.49\textwidth}
    \begin{center}
    \subfigure[ ]{
        \includegraphics[width=1.\textwidth]{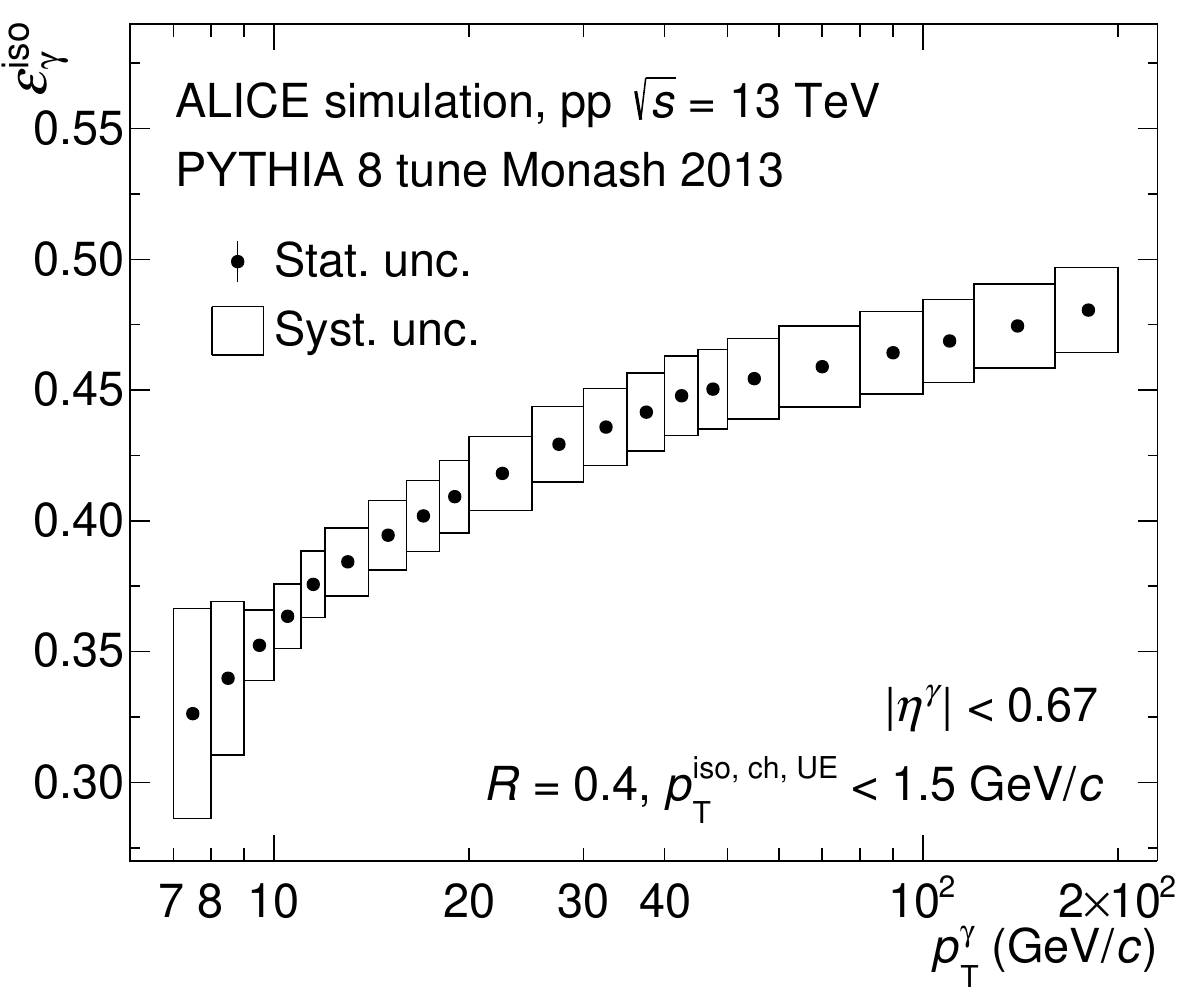}
        \label{fig:effiPub}
    }
    \end{center}
    \end{minipage}
\end{center}
\caption{\small {  (colour online) 
The different contributions (reconstruction, identification, isolation) and the total isolated-photon efficiency as a function of the reconstructed \ptg\ are shown in panel (a) and (b), respectively. The systematic uncertainty shown as boxes in panel (b) have been obtained from the ``no MC tuning” uncertainty source discussed in Sect.~\ref{sec:sys_unc}.
}}
\label{fig:EffIsoPhoton}
\end{figure}

The different analysis selection criteria determine the overall efficiency and the contributions are presented in Fig.~\ref{fig:effiComponent}.
They are calculated as the ratio of spectra, where the denominator is the number of generated photons ${\rm d}N ^{\rm gen}_{\gamma} / {\rm d}p_{\rm T}^{\rm gen}$, and the factors in the numerator are the reconstructed spectra after different selection criteria, ${\rm d}
N ^{\rm rec}_{\rm cut} / {\rm d}p_{\rm T}^{\rm rec}$. 
The different contributions are the following:
(i) the pure reconstruction efficiency of photons is $\varepsilon^{\mathrm{rec}} \approx 50 \%$, (green squares), where the efficiency loss is mainly due to excluded regions in the calorimeter, in particular, the requirement of $d_{\rm mask}>2$, and exclusion of 
clusters close to the border of EMCal supermodules;
(ii) the photon identification (shower shape selection) reduces the efficiency by 10--20\%, leading to $\varepsilon^{\mathrm{rec}} \times \varepsilon^{\mathrm{id}} \approx 35-45 \%$, (red crosses); 
(iii)  the isolation criterion decreases the efficiency to $\varepsilon^{\mathrm{rec}} \times \varepsilon^{\mathrm{id}} \times \varepsilon^{\mathrm{iso}} \approx 30-40$\%, (blue diamonds). 
The efficiency is \pt\ dependent due to the \sigmalong\ selection
because the photon peak is wider at lower \pt. This is already present in $\varepsilon^{\mathrm{rec}}$ due to the selection \sigmalong~$> 0.1$, which is applied to reject anomalous energy depositions;
In addition, (iv) the fraction $\kappa^{\mathrm{iso}}$ of generated photons which are isolated is represented by black-filled circles in Fig.~\ref{fig:effiComponent}. The total efficiency corresponds to the ratio of the reconstruction, identification and isolation efficiency (iii) to the isolated generated photon fraction (iv) and is then directly calculated as follows
\begin{equation}
\label{eq:efficiency}
     \varepsilon_{\gamma}^{\rm iso} = \frac{{\rm d}N ^{\rm rec}_{\rm n,\,iso}} { {\rm d}p_{\rm T}^{\rm rec}} {\Bigg /} \frac{{\rm d}N ^{\rm gen}_{\gamma,\,\rm iso}} { {\rm d}p_{\rm T}^{\rm gen}}
     \equiv \frac{\varepsilon^{\mathrm{rec}} \times \varepsilon^{\mathrm{id}} \times \varepsilon^{\mathrm{iso}}}{\kappa^{\mathrm{iso}}},
\end{equation}
where $N^{\rm rec}_{\rm n,~iso} $ is the number of 
clusters which are reconstructed and identified as isolated photons and which are produced by a prompt photon,  
and $N^{\rm gen}_{\gamma,~\rm iso}$ is the number of generated prompt photons 
which pass the isolation momentum threshold in the same way as at the detector level.
The overall efficiency for the reconstruction of isolated photons ranges from approximately 30 to 45\% as shown in  Fig.~\ref{fig:effiPub}. 

\subsection{Trigger efficiency, rejection factor and luminosity}

The isolated-photon yield correction needs to take into account the performance of the calorimeter trigger in particular when calculating the event normalisation and luminosity.

The EMCal L1-$\gamma$-low and -high trigger efficiency $\varepsilon_{\rm trig}$ is the probability that the trigger selects events when a high-energy cluster is reconstructed in the EMCal acceptance above a given trigger threshold. 
The trigger efficiency is however not 100\% above the trigger threshold because of reduced geometric coverage of the trigger compared to the EMCal acceptance due to trigger cell tiles ($2\times 2$ cells) and full TRU cards (Trigger Region Units, $24\times16$ cells in $\varphi \times \eta$) that were inactive or masked. It is also \pt\ dependent since the cluster can cover more cells the higher the energy (owing to nearby jet particles in the event and meson decay merging), being less affected by small masked regions.

\begin{figure}[ht]
\begin{center}
    \begin{minipage}{0.483\textwidth}
    \begin{center}
    \subfigure[ ]{
        \includegraphics[width=1\textwidth]{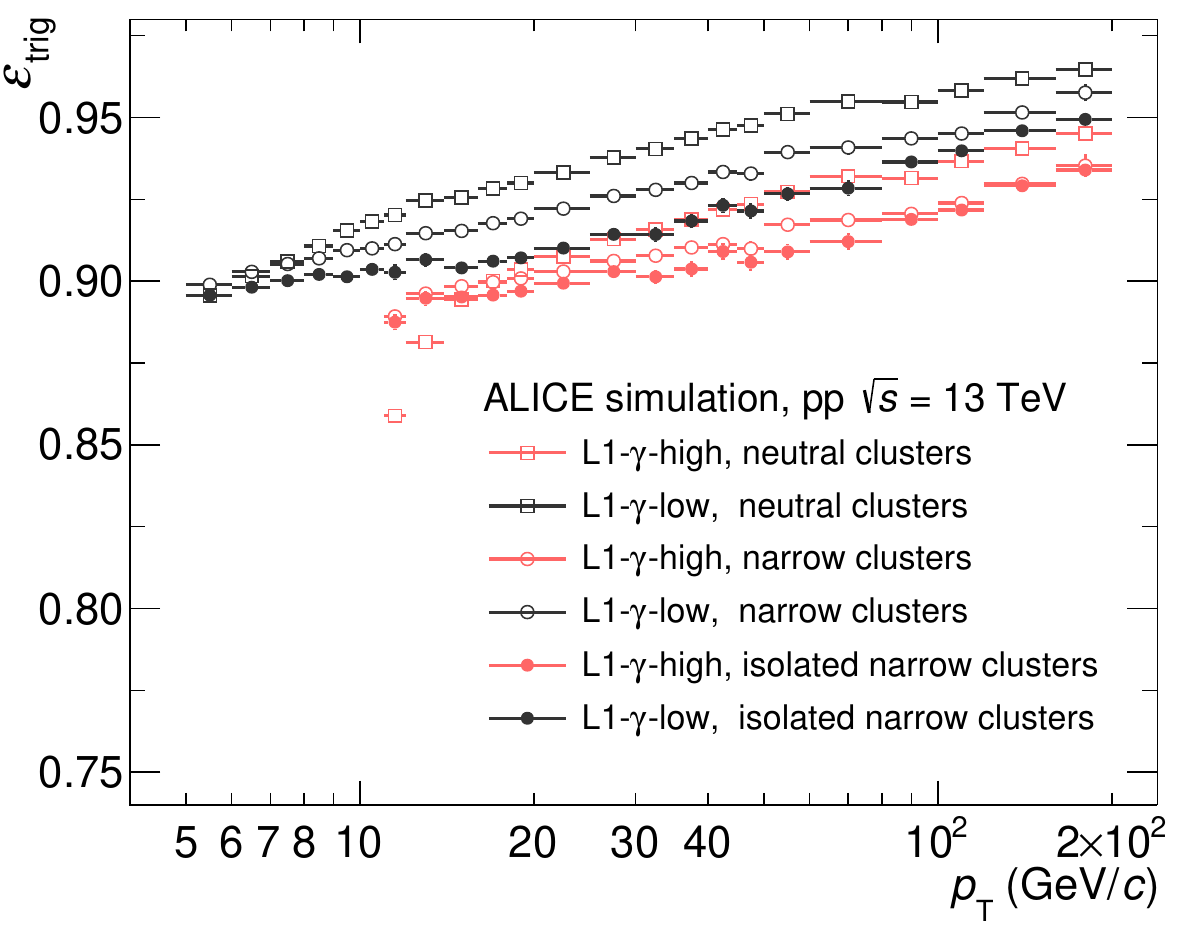}
        \label{fig:effiTrig}
    }
    \end{center}
    \end{minipage}
    \begin{minipage}{0.485\textwidth}
    \begin{center}
    \subfigure[ ]{
        \includegraphics[width=1\textwidth]{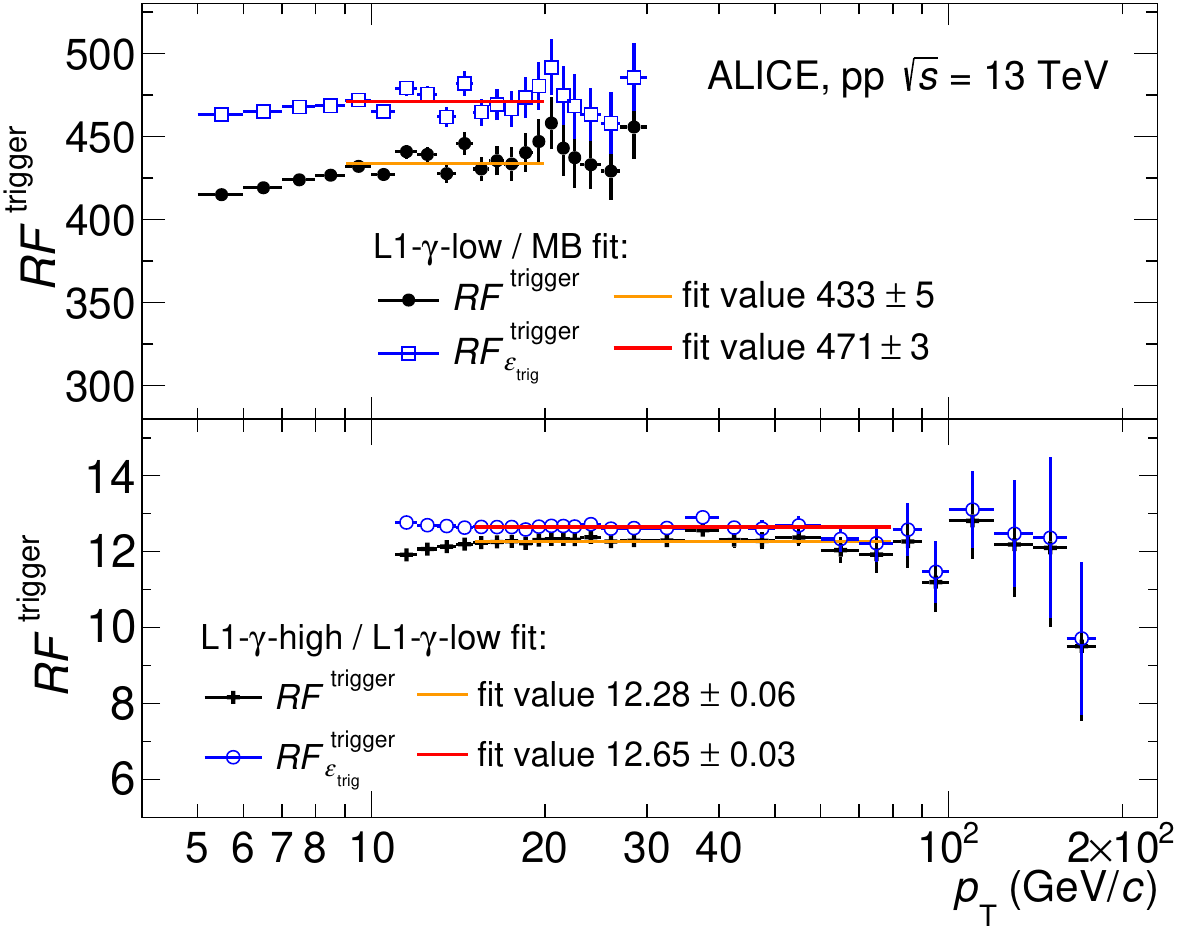}
        \label{fig:RF}
    }
    \end{center}
    \end{minipage}
\end{center}
\caption{\small {  (colour online) 
(a) Trigger efficiency for neutral clusters, narrow clusters, and isolated narrow clusters for the L1-$\gamma$-low and -high calorimeter triggers. (b) Trigger rejection factor calculated with and without applying the trigger efficiency for each of the calorimeter triggers. 
The results of a constant fit to the rejection factor plateau regions are shown in the legend, while the estimation of the uncertainties are described in Sect.~\ref{sec:sys_unc}.
}}
\label{fig:EffTrig_RF}
\end{figure}

The trigger efficiency is calculated from simulation, combining the jet--jet and $\gamma$--jet simulations, applying the same trigger logic as in the data, and it is shown in Fig.~\ref{fig:effiTrig}. The trigger efficiency for neutral clusters ($\varepsilon^{\rm clus}_{\rm trig}$) and for the lower threshold varies from close to 90\% at cluster $\pt = 5$~\GeVc to close to 97\% at 200~\GeVc. For the higher threshold, the efficiency is lower, close to 88\% at 12 GeV/$c$ and 95\% at 200~\GeV/c.
The narrow clusters have a lower efficiency for firing the trigger 
because they are more affected by cell masking than wide clusters. In addition, the isolation tends to select even narrower clusters and, therefore, the trigger efficiency for narrow clusters after isolation ($\varepsilon^{\rm iso}_{\rm trig}$) is a few \% lower when moving 
significantly away from the trigger threshold. 
The behaviour close to the threshold is reversed between neutral and narrow clusters, most likely because very wide clusters are less efficient in passing the threshold in a trigger tile.

The EMCal trigger rejection factor, $RF^{\rm trigger}$, quantifies the enhancement fraction of calorimeter triggers with respect to MB triggers. 
It is calculated via the ratio of the calorimeter neutral-cluster \pt spectra
\begin{equation}
RF^{\rm trigger}=\frac{1/N_{\rm evt}^{\rm L1\text{-}\gamma}\times {\rm d}N^{\rm L1\text{-}\gamma}/{\rm d}p_{T}} { 1/N_{\rm evt}^{\rm MB}\times {\rm d}N^{\rm MB}/{\rm d}p_{T}},
\end{equation}
where $N_{\rm evt}^{\rm trigger}$ is the number of events for a given trigger, as reported in Table~\ref{tab:NevtRFLumi}.  
This ratio increases quickly with increasing \pt\ below the trigger threshold and reaches a plateau slightly above the threshold. The plateau is fitted to a constant and the result gives the trigger enhancement. 
Note that the plateau is not completely \pt\ independent 
as observed also in the trigger efficiency, which leads to a not well-constrained fit.
To correct for that \pt\ dependence and have a better fit result, 
the triggered cluster spectrum was corrected by
the neutral-cluster trigger efficiency
\begin{equation}
RF^{\rm trigger}_{\varepsilon_{\rm trig}}=\frac{1/N_{\rm evt}^{\rm L1\text{-}\gamma} \times {\rm d}N^{\rm L1\text{-}\gamma}/{\rm d}p_{T} \times 1/{\varepsilon_{\rm trig}^{\rm clus}}} { 1/N_{\rm evt}^{\rm MB} \times {\rm d}N^{\rm MB}/{\rm d}p_{T}} = \frac{RF^{\rm trigger} }{\varepsilon_{\rm trig}^{\rm clus}}.
\end{equation}
Figure~\ref{fig:RF} shows both trigger rejection factors 
when comparing the MB trigger to the low threshold trigger, $RF^{\rm L1\text{-}low,~MB}$ and the latter one to the high threshold, $RF^{\rm L1\text{-}high,~L1\text{-}low}$.
In $RF^{\rm L1\text{-}low,~MB}$, 
one can observe that the plateau is flatter near the trigger threshold below 10~\GeVc when the trigger efficiency is used 
in the calculation. This is not observed in $RF^{\rm L1\text{-}high,~L1\text{-}low}$ for which the \pt dependence is similar. 
Since the trigger rejection factor is corrected by the trigger efficiency, 
which increases $RF^{\rm L1\text{-}low,~MB}$ by about 10--12\%,
the trigger efficiency for isolated photons must also be used to correct the final isolated-photon yield, otherwise, the final yield would be artificially enhanced.

The integrated luminosity collected with each of the used triggers 
($\mathscr{L}_{\rm int}^{\rm trigger}$) 
has been determined using the expression
\begin{equation}
\mathscr{L}_{\rm int}^{\rm trigger} = \frac{N_{\rm evt}^{\rm trigger} \times RF^{\rm trigger}_{\varepsilon_{\rm trig}}} { \sigma_{\rm MB}} 
\end{equation}
where $\sigma_{\rm MB}=58.05 \pm 0.90$~mb~\cite{ALICE-PUBLIC-2016-002,ALICE-PUBLIC-2021-005} is the measured minimum-bias trigger cross section 
averaged over the three data-taking years.
The corresponding values of the $RF^{\rm trigger}_{\varepsilon_{\rm trig}}$ and integrated luminosity per trigger are presented in Table~\ref{tab:NevtRFLumi}. 

\begin{table}[htbp]
\begin{center}
\caption{\label{tab:NevtRFLumi} Number of selected events, EMCal L1-$\gamma$ trigger rejection factors, and luminosity per trigger. 
The luminosity uncertainty contains both the $\sigma_{\rm MB}$ and rejection factor uncertainties.}

\setlength{\tabcolsep}{7mm}{
\begin{tabular}{ c l c c c}
 Trigger             & $N_{\rm evt}^{\rm trigger}$ & $RF^{\rm trigger}_{\varepsilon_{\rm trig}}$ & $\mathscr{L}_{\rm int}^{\rm trigger}$\\
\hline
 MB                  & $1.587 \times 10^{9}$ & ~             &  27.34 $\pm$ 0.42 nb$^{-1}$ &\\
 L1-$\gamma$-low     & $1.356 \times 10^{8}$ & 471 $\pm$ 3   &   1.13 $\pm$ 0.02 pb$^{-1}$ &\\
 L1-$\gamma$-high    & $9.354 \times 10^{7}$ & 5960 $\pm$ 40 &   9.63 $\pm$ 0.16 pb$^{-1}$ &\\
\hline

\end{tabular}}

\end{center}
\end{table}

The final production cross section is measured as a function of \pt, thus, the different triggers are combined depending on the trigger threshold. The L1-$\gamma$-high trigger threshold is at $E=9$~GeV but it is not fully efficient until few GeV after, so it was decided to combine the three triggers above $\pt=12$\,\GeVc. 
Hence, the distribution below $\pt=12$\,\GeVc is a combination of MB and L1-$\gamma$-low triggers. 
The resulting sampled luminosity of the current measurement is calculated as

\begin{equation}
\begin{split}
\mathcal{L}_{\rm int} (\pt > 12~\mathrm{GeV/}c) = \mathcal{L}_{\rm int}^{\rm MB} + \mathcal{L}_{\rm int}^{\rm L1\text{-}\gamma\text{-}low} + \mathcal{L}_{\rm int}^{\rm L1\text{-}\gamma\text{-}high}~\mathrm{and} ~\\                  
\mathcal{L}_{\rm int} (\pt \leq 12~\mathrm{GeV/}c) = \mathcal{L}_{\rm int}^{\rm MB} + \mathcal{L}_{\rm int}^{\rm L1\text{-}\gamma\text{-}low},~~~~~~~~~~~~~~~~~~~~~~~~~~~
\end{split}                     
\label{eq:lint}
\end{equation}
resulting in  
\begin{equation}
\begin{split}
\mathscr{L}_{\rm int} (\pt > 12~\mathrm{GeV/}c)=10.79
\pm 0.16~\mbox{pb}^{-1},~~\\
\mathscr{L}_{\rm int} (\pt \leq 12~\mathrm{GeV/}c)=1.160\pm 0.018~\mbox{pb}^{-1}.
\end{split}
\label{eq:lintTot}
\end{equation}

\clearpage
\section{Systematic uncertainties}
\label{sec:sys_unc}

\begin{figure}[ht]
    \begin{minipage}{0.49\textwidth}
    \centering
        \subfigure[ ]{
        \includegraphics[width=1.\textwidth]{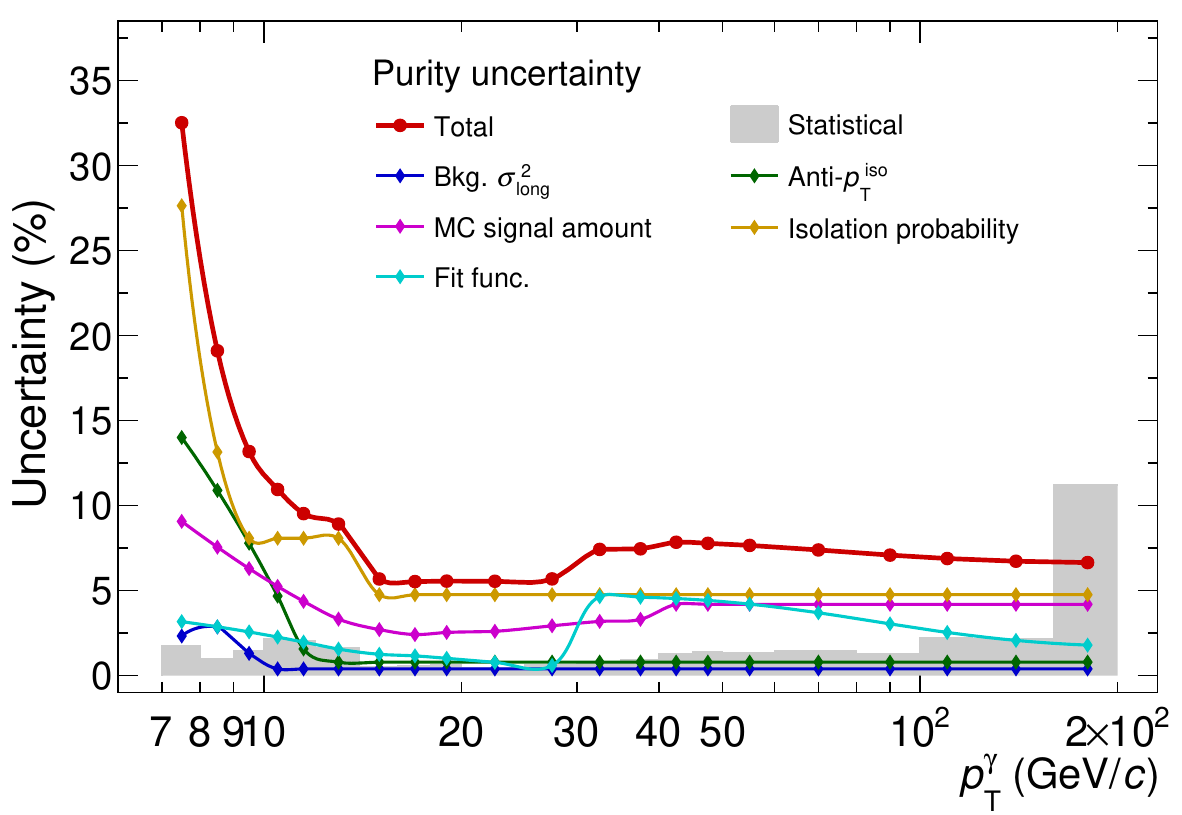}
        }
    \end{minipage}
    \begin{minipage}{0.49\textwidth}
    \centering
        \subfigure[ ]{
        \includegraphics[width=1.\textwidth]{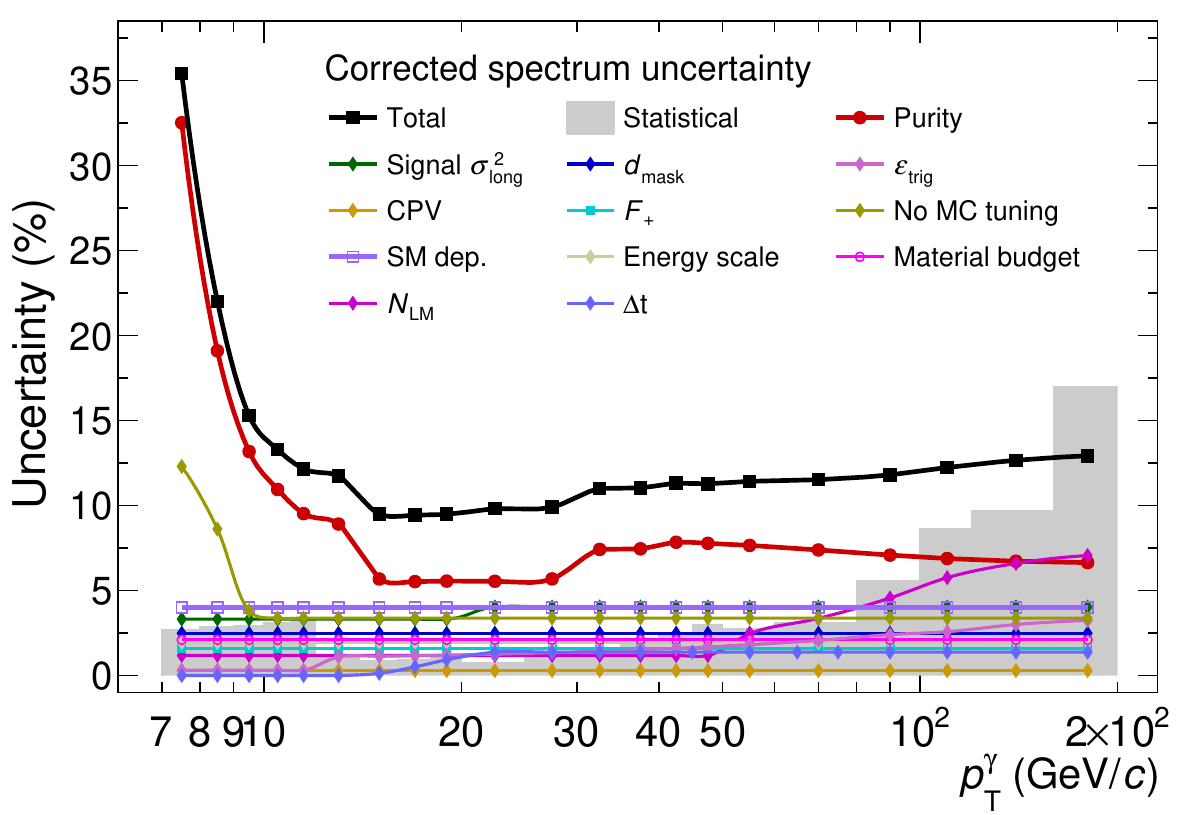}
        }
    \end{minipage}
    \caption{\label{fig:sys_unc} (colour online) 
    Relative systematic uncertainty sources of the isolated-photon purity (a) and cross section yield (b) and their quadratic sum as a function of \ptg. The statistical uncertainty is also shown for reference as a shaded histogram. The total purity uncertainty in (a) is one of the sources added in quadrature to the total cross section uncertainty in (b). 
    }
\end{figure}

Figure~\ref{fig:sys_unc} displays the estimated relative systematic uncertainties as a function of \ptg\ for all the considered sources.
The uncertainty contributions from all the sources are added in quadrature.
The contributions that enter into the purity calculation and those that enter into the yield are shown separately, and the latter also includes the total uncertainty for the purity.
The values of the systematic uncertainties for the different sources are reported in Table~\ref{tab:systsummary} for two extreme transverse momentum intervals for the two combinations of triggers used in the analysis.

\begin{table}[hb]
\centering
   \caption{
   Summary of uncorrelated relative systematic uncertainties in per cent for selected \ptg intervals of the isolated-photon measurement.
   The purity uncertainty is included in the yield total uncertainty.
   The statistical uncertainty is also shown for reference.
   The luminosity normalisation uncertainty of 1.5\% from Eq.~\eqref{eq:lintTot} is not included in this table.
   }
   \label{tab:systsummary}
   \begin{tabular}{l l l l l }
   \hline
   $\ptg$ (\GeVc)                 & 7--8     & 10--11  & 40--45  & 160--200 \\
   \hline
   MC signal amount              & 9.1\%   & 5.2\%   & 4.2\%   & 4.2\% \\
   \sigmalong\ background range  & 2.3\%   & 0.4\%   & 0.4\%   & 0.4\% \\
   Anti-\ptIso\ range            & 14.0\%  & 4.7\%   & 0.8\%   & 0.8\% \\
   Isolation probability         & 27.6\%  & 8.1\%   & 4.8\%   & 4.8\% \\
   Fit function                  & 3.2\%   & 2.3\%   & 4.5\%   & 1.8\% \\
   \hline
   Purity total unc.             & 32.5\%  & 10.9\%  & 7.8\%   & 6.6\% \\
   \hline
   Charged particle veto         & 0.3\%   & 0.3\%   & 0.3\%    & 0.3\% \\
   $N_{\rm LM}$                  & 1.2\%   & 1.2\%   & 1.2\%    & 7.1\% \\
   $d_{\rm mask}$                & 2.5\%   & 2.5\%   & 2.5\%    & 2.5\% \\
   $F_{+}$                       & 1.6\%   & 1.6\%   & 1.6\%    & 1.6\% \\
   $\Delta t$                    & 0\%     & 0\%     & 1.4\%    & 1.4\% \\
   \sigmalong\ signal range      & 3.3\%   & 3.3\%   & 4.0\%   & 4.0\% \\
   No MC tuning  ($\varepsilon_{\gamma}^{\rm iso}$ unc.) 
                                 & 12.3\%  & 3.4\%   & 3.4\%   & 3.4\% \\
   Trigger effic. $\varepsilon_{\rm trig}$
                                 & 0.3\%   & 0.3\%   & 1.6\%    & 3.2\% \\
   Energy scale                  & 2.0\%   & 2.0\%   & 2.0\%   & 2.0\% \\ 
   Material budget               & 2.1\%   & 2.1\%   & 2.1\%    & 2.1\% \\
   SM dependence                 & 4.0\%   & 4.0\%   & 4.0\%    & 4.0\% \\
   \hline
   Yield total unc.              & 35.4\%  & 13.3\%  & 11.3\%   & 12.9\%\\
   Statistical unc.\             & 2.7\%   & 3.1\%   & 2.4\%    & 17.0\%\\
   \hline
   \end{tabular}
\end{table}

The uncertainty contributions assigned to the purity correction are estimated from variations of the isolation momentum background ranges, shower shape background ranges, isolation probability, signal amount in the simulation, and errors of the fit to the purity.

The amount of signal in the simulation, labelled as the ``MC signal amount" in the table and figure, influences the aforementioned leakage effect of signal into the background regions used to estimate the purity.
This is checked using different weights assigned to the signal in the simulation  ($\gamma$--jet PYTHIA~8 events), here $\pm 50$\%.
The resulting uncertainty is 9.1\% at 7~\GeVc decreasing to 2.5\% at 20~\GeVc 
and then increasing up to 40~\GeVc beyond which it remains constant at 4.2\%. 

The uncertainty due to the choice of the background region range (wide showers, i.e.\ large values of \sigmalong) is investigated by moving the corresponding \sigmalong\ interval to $ 0.37 <$~\sigmalong~$ <2.37$ and $ 0.43 <$~\sigmalong~$<2.43$. The estimated uncertainty is found to be 1--3\% below 10~\GeVc\ and 0.4\% above.
The \ptIso\ background range, labelled as the ``anti-\ptIso" in the figure and table, is also varied, with the minimum limit from 2 to 5~\GeVc\ and the maximum from 6 to 20~\GeVc. 
The average of the differences due to these variations is used
to estimate the uncertainty, resulting in an uncertainty of 14\% at 7~\GeVc decreasing to 0.8\% at 12~\GeVc from where it remains constant.
 
 The systematic uncertainty related to the correlation effects 
 between \ptIso\ and \sigmalong\ discussed in the Sect.~\ref{sec:purity}  
 is labelled as ``isolation probability”.  The uncertainty is obtained
 by the difference between the \alphaf\ factors obtained without and with the calibration according to Eq.~\eqref{eq:CorrectionMCExtra}, changing the \sigmalong\ background region fit range variation. The average of the differences is used as uncertainty.
 The resulting uncertainty is estimated to decrease from 27.6\% at $\ptg = 7$\,\GeVc to 8.1\% at $\ptg = 9$--14\,\GeVc, and  to a constant 4.8\% for $\ptg > 14$\,\GeVc. 

\newpage

The fit of the purity with sigmoid functions has an associated uncertainty based on the parameter fit uncertainties presented in the previous section. The three parameters of the function were varied within their uncertainties, and the maximum deviation of the function is used as uncertainty. For the low-\ptg part of the fitting function, the uncertainty decreases from 3\% to 0.5\% from 7 to 30~\GeVc, and for the high-\ptg part, the uncertainty decreases from 4.5\% to 2\% from 30 to 200~\GeVc. 

The contributions from all these uncertainty sources are added in quadrature and are used as total purity uncertainty that ranges between 6\% and 35\% being maximal at low \ptg\ and having a minimum for \ptg\ between 16 and 30~\GeVc.
The main source of systematic uncertainty is the isolation probability. The second most important source of uncertainty is \ptg-dependent: at low \ptg\ (up to 10~\GeVc) the anti-isolation range, between 30 and 60~\GeVc the uncertainty of the fit to the purity, and for the other \ptg-ranges the MC signal amount.

The uncertainties in the cross section yield due to the choice of the neutral cluster selection criteria in this analysis are evaluated via the variations of the charged particle veto residual distance selection, the number of local maxima, distance to masked channels, and the parameter $F_{+}$. 
The uncertainty due to the charged particle veto is estimated by varying the parameters of the track \pt-dependent selection of Eq.~\eqref{eq:cpv} to looser ones $\Delta \eta^{\rm residual}>0.025$ and $\Delta \varphi^{\rm residual}>0.03 $ radians. 
The resulting uncertainty on the cross section is constant with $\ptg$ at 0.3\%.
The uncertainty related to the selection on the number of local maxima is obtained by varying the threshold from 2 to 3 maxima, and it ranges from 1.2--2\% for \ptg$<50$~\GeVc\ to 8\% at the highest \ptg.
The requirement on the distance to masked channels is decreased from $d_{\rm mask}>2$ to 0 cells, 0 cells meaning that the misbehaving or dead channels are effectively removed from the cluster but the presence of these cells in the proximity of the cluster has no further effect, and a constant uncertainty of 2.5\% is obtained. 
The $F_{+}$ is varied from 97\% to 95\% and a constant variation of 1.6\% is observed. 
The time selection window was varied between $\Delta t = 10$ and 40~ns to study the effect of pileup and cells with anomalous depositions that pass the $F_{+}$ selection, 
the uncertainty was found to be negligible below 16~\GeVc and it increases to 1.4\% for $\ptg>20$~\GeVc. 

The choice of the signal range of the \sigmalong\ of narrow photon-like showers is important for the efficiency calculation but also influences the background estimate via a “leakage” of photon showers to the control regions. The uncertainty due to the choice of the signal range is estimated by varying the upper limit of the range to \sigmalong~=~0.27 and 0.37 and is found to lie at 3.3\% below \ptg = 20~\GeVc and 4\% above.

The description of the shower shape in simulations can also affect the efficiency measurement, while the effect on the purity is found to be negligible. 
The associated uncertainty decreases with \ptg\ from 12\% at 7~\GeVc to 3.4\% at 10~\GeVc and above. It is estimated from the difference between standard simulations and those 
including modelling of the cross talk observed in the EMCal readout cards and is labelled as ``no MC tuning" in the table and figure. 

The uncertainty on the trigger normalisation has two sources: the use of the trigger efficiency to estimate the trigger rejection factor and correct the yields, and the fit used to calculate the trigger rejection factor. For the first source, 
the comparison of the yields calculated with either using or not the trigger efficiency was considered, 
and half of the difference is taken. The uncertainty amounts to 0.25\% when ${\rm L1\text{-}}\gamma{\rm \text{-}low}~+~{\rm MB}$ triggers are used and 1 to 3\% from 12 to 200~\GeVc when all the triggers are combined.  

The trigger rejection factor for the lower L1-$\gamma$ trigger threshold with respect to the MB trigger, denoted as  $RF^{\rm L1\text{-}low,~MB}_{\varepsilon_{\rm trig}}$, and for the higher L1-$\gamma$ 
 trigger threshold with respect to the lower L1-$\gamma$ trigger threshold, denoted as $RF^{\rm L1\text{-}high,~L1\text{-}low}$, is calculated fitting with a constant above the trigger threshold when it is fully efficient. The fitting range is varied for $RF^{\rm L1\text{-}low,~MB}_{\varepsilon_{\rm trig}}$, and the calculated standard deviation of all the variations gave a 0.2\% uncertainty. A similar procedure was applied for $RF^{\rm L1\text{-}high,~L1\text{-}low}_{\varepsilon_{\rm trig}}$ resulting in an uncertainty of 0.6\%. The rejection factor for the L1-$\gamma$-high trigger is obtained by multiplication of the previous two, its uncertainty is estimated as the quadratic sum of their respective uncertainties corresponding to 0.7\%.
This uncertainty is considered as a normalisation uncertainty and not added to the yield systematic uncertainty. These uncertainties combined with the $\sigma_{\rm MB}$ uncertainties give a normalisation and luminosity uncertainty of 1.5\% for both trigger \ptg ranges considered from Eq.~\eqref{eq:lintTot}.

The uncertainty on the energy scale of the EMCal is estimated to be 0.5\%~\cite{ALICE:2022qhn}.
The effect of this uncertainty on the measured cross section amounts to 2\%.
A material budget uncertainty accounting 
for an imperfect description in the simulation of the material
of the different detectors traversed by photons before they reach the EMCal has been previously determined in Ref.~\cite{ALICE:2018mjj} and amounts to 2.1\%. 

Due to the different hardware and electronics performance of the SMs, the result can potentially change depending on the SM where the cluster is measured. To estimate the effect, the neutral cluster yield measured in each SM in data is divided by the yield in simulation, and each of these yield ratios is divided by the full calorimeter neutral cluster yield in data over simulation. In ideal conditions, this double ratio should be equal to unity and \pt-independent. The double ratios obtained are \pt-independent and close to unity, although with small deviations. The dispersion of those double ratios is found to be 4\%, and this value is assigned as the SM-dependent uncertainty.

The total systematic uncertainty on the cross section is obtained by adding in quadrature the contributions of the different sources described above, as well as the purity uncertainty. 
The resulting uncertainty decreases from close to 35\% at $\ptg = 7$~\GeVc to close to 10\% at $\ptg =25$~\GeVc  and it increases slowly at higher \ptg\ reaching close to 13\% at $\ptg =200$~\GeVc. 
The dominant source of the systematic uncertainty is the total purity uncertainty up to $\ptg =100$~\GeVc, while for larger \ptg\ the effect of varying the selection on the number of local maxima becomes dominant. 
Nonetheless, for $\ptg =100$~\GeVc, the measurement is dominated by statistical uncertainties.

\clearpage
\section{Results\label{sec:results}}

The isolated-photon production differential cross section can be obtained from the following equation

\begin{equation}
\label{eq:cs}
\frac{{\rm d}^{2}\sigma_{\gamma}^{\rm iso}}{{\rm d}\ptg~{\rm d}\eta} =  \frac{\sigma_{\rm MB}}{\sum_{i} N_{\rm evt}^{i} \times {RF}^{i}} \left(\sum_{i} \frac{{\rm d}^{2}N_{\rm n}^{{\rm iso},~i}}{{\rm d}\ptg~{\rm d}\eta} \times \frac{1}{\varepsilon_{\rm trig}^{{\rm iso},~i}} \right)\times \frac{P}{\varepsilon_{\gamma}^{\rm iso} \times \kappa^{\rm iso} \times \text{Acc}}
\end{equation}

where all the terms were described in the previous sections and $i$ is an index depending on the data trigger and Acc~$=\Delta \eta \times \Delta \varphi / 2\pi$ is the acceptance area of the analysis obtained from the values in Table~\ref{tab:acceptance}. The cross section formula includes a $\kappa^{\rm iso}$ dividing term (see definition in Sect.~\ref{sec:efficiency}) to take into account the bias induced by the collision underlying event not present in NLO calculations. This factor is calculated with PYTHIA~8 $\gamma$--jet events at the generator level. It is found that multi-particle interaction processes cause most of the deviation from unity, and only at the highest \ptg\ a small contribution from initial and final state 
radiation effects induces a deviation from unity by up to 3\%.
In the measurement at $\s = 7$~\TeV~\cite{ALICE:2019rtd}, this correction was multiplied to the NLO calculations instead of the data, this time it is considered more appropriate to correct the measurement so that comparisons with theory and other \s\ are simplified. Note that if the UE is subtracted from the isolation cone, this correction is not needed.

\begin{figure}[tb]
    \begin{center}
    \includegraphics[width=0.8\textwidth] {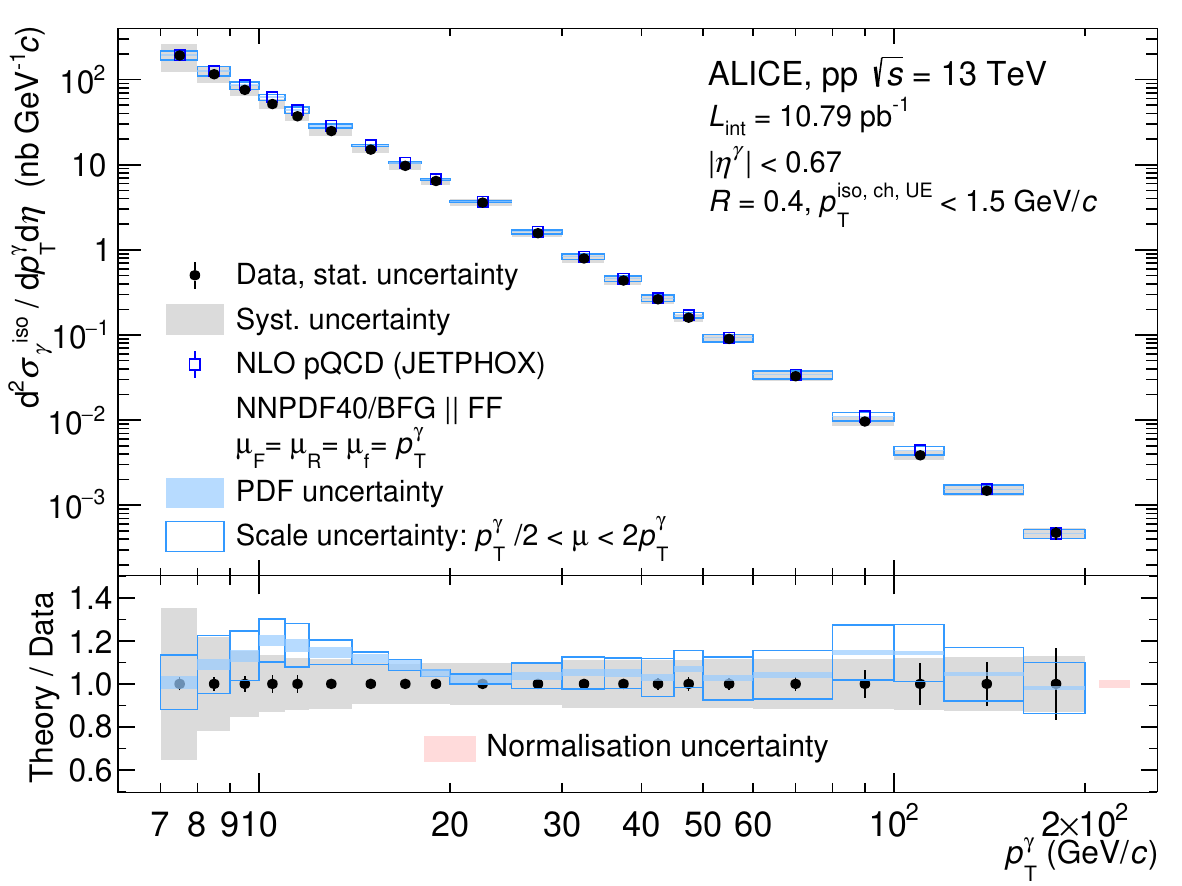}
    \end{center}
    \caption{\label{fig:isoPhotonCrossSection} (colour online)  
    Top panel: Differential cross section of isolated photons measured in pp collisions at $\sqrt s =13$~TeV. Vertical black lines and grey-filled boxes represent data statistical and systematic uncertainties, respectively. The blue boxes correspond to pQCD calculations with JETPHOX, open boxes for scale uncertainty and filled boxes for PDF uncertainty, respectively. Bottom panel: Ratio between the JETPHOX calculation results and data displayed in the blue boxes, vertical boxes size shows the theory scale and PDF uncertainties. Experimental uncertainties are shown here on the black points centred at unity. The normalisation uncertainty of 1.5\%, described in the text, is included only in the bottom panel and shown as a red box at 200~\GeVc.  
}
\end{figure}

Figure~\ref{fig:isoPhotonCrossSection} shows in the top panel the isolated-photon cross section as a function of \ptg and the theory over data ratio in the bottom panel. Error bars indicate the statistical uncertainties and boxes the systematic uncertainties, respectively. 
An additional normalisation uncertainty of 1.5\% coming from both measured minimum bias cross section and the rejection factors from the EMCal triggering from Eq.~\eqref{eq:lintTot} is not added to the systematic uncertainties
on the data points, but shown as a separate box in the bottom panel with the theory-to-data ratio.

The measurement is compared to NLO pQCD calculations using JETPHOX 1.3.1~\cite{Fontannaz, Aurenche}. The parton distribution function used in the calculations is  NNPDF4.0~\cite{Ball_2022}, and the fragmentation function is BFG~II~\cite{PFFth}. The central values of the predictions were obtained by choosing factorisation, normalisation, and fragmentation scales equal to the photon transverse momentum ($\mu_{f}=\mu_{R}=\mu_{F}=\ptg$). Scale uncertainties were determined varying all scales simultaneously to 0.5 and 2 times their nominal values.
The uncertainties related to the PDF are obtained by performing calculations for each of the 101 members of NNPDF4.0. The resulting uncertainties are reported at 90\% CL. 
The isolation criterion in pQCD calculations corresponds to a restriction of the available phase space to final-state radiation in a cone of $R < 0.4$~\cite{Fontannaz}.
The threshold of $p_{\rm T}^{\rm iso} < $~2 \GeVc, which includes both charged and neutral particle transverse momenta, is used since it is equivalent to the one in data of $\ptIso < 1.5$\,\GeVc that only uses transverse momenta of charged particles. 
The theoretical predictions describe the measured isolated-photon cross section within uncertainties in the full transverse momentum range of the measurement as demonstrated by the theory-to-data ratio shown in Fig.~\ref{fig:isoPhotonCrossSection} (bottom). 
In particular, above 20~\GeVc the deviations of the ratio from unity are small, a constant fit to the points gives 3.6\%.
The measurement is done for the first time in the range 7--10~\GeVc in any collision system at the LHC, showing agreement with the theory.
The ratio for $10<\ptg<20$~\GeVc tends to deviate from unity, although it is still in agreement within uncertainties. 
The small deviation between the data and the theory predictions in this \ptg\ region might in part be due to a small potential under-correction of the purity, which has a steep rise in the 10--20~\GeVc interval. Note also that the systematic uncertainty on the measured cross section in this \ptg\ interval is significantly dominated by the uncertainty on the purity determination.

\begin{figure}[ht]
    \centering
    \includegraphics[width=0.8\textwidth]{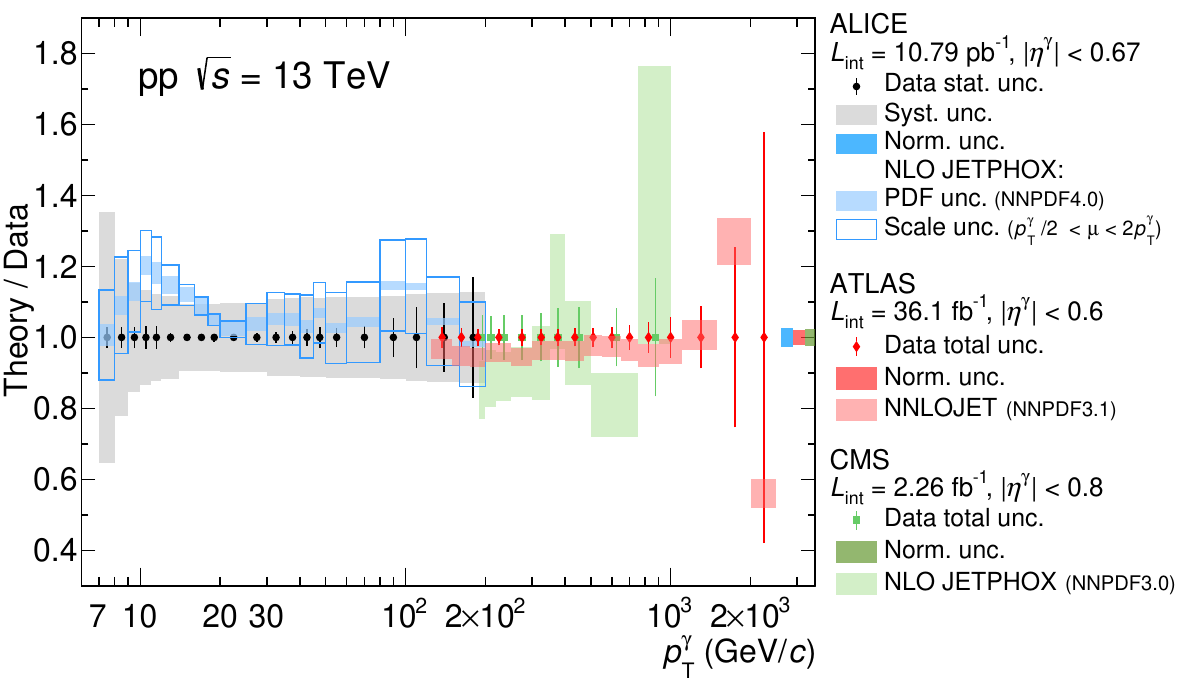}
    \caption{\label{fig:isoPhotonCrossSectionComp}(colour online) Ratio between
    theory predictions for the isolated-photon differential cross section and
    measurements for ATLAS~\cite{Aad:2019}, CMS~\cite{Sirunyan2019}, and ALICE. Theory predictions are obtained for ATLAS with NNLOJET~\cite{Chen_2020} NNLO NNPDF3.1 PDFs~\cite{Ball_2017}, for CMS with JETPHOX NLO and NNPDF3.0 PDFs~\cite{Ball_2015}, and for ALICE with JETPHOX NLO and NNPDF4.0 PDFs~\cite{Ball_2022}. Experimental uncertainties are shown here on the points centred at unity, for CMS and ATLAS statistical and systematic uncertainties are added in quadrature and shown as a vertical error bar. ALICE error bars are statistical uncertainties and boxes are systematic uncertainties. The coloured boxes centred at the ratio indicate the theoretical uncertainty on the PDF and scales. 
    The luminosity normalisation uncertainty of each experiment is presented as an overall box around unity at the right of the figure centred at unity. 
    }
\end{figure}

Figure~\ref{fig:isoPhotonCrossSectionComp} shows the comparison of the ratios between the theory predictions and the measurements for the \pt-differential isolated-photon cross sections from three different LHC experiments, namely ALICE (NLO), ATLAS~\cite{Aad:2019} (NNLO), and CMS~\cite{Sirunyan2019} (NLO). 
The comparison is made on ratios of data to prediction using equivalent isolation criteria since those criteria differ
among these experiments such that a direct comparison of the isolated-photon cross sections is not fully adequate.
There is a small overlap region between ALICE and the other experiments in the \ptg\ interval between 125 (ATLAS) and 190 (CMS) to 200~GeV/$c$, 
where the ratios are in agreement within the uncertainties.
It is worth noticing the contribution of ALICE to the isolated-photon measurements in pp collisions at \s~$=13$~TeV since the \ptg\ range is decreased by an order of magnitude compared to ATLAS and CMS.
The ATLAS and CMS results use larger values and different definitions for the isolation momentum selection criterion, about 5~\GeVc\ for CMS and approximately 4~\GeVc\ at $\ptg=200$~\GeVc and increasing with \ptg for ATLAS,
both with the same cone radius $R=0.4$ but 
including contributions from both neutral and charged particles.
In JETPHOX predictions, increasing the isolation threshold should reflect a larger fragmentation contribution in the total cross section without necessarily increasing the total isolated-photon cross section compared to smaller isolation momentum selection criteria. 
However, the theory-to-data ratios should be consistent between the experiments as is observed in the figure. 
The ALICE NLO-to-data ratio for $\ptg > 20$~\GeVc is almost on top of unity, while CMS (NLO) and ATLAS (NNLO) ratios are consistently below but in agreement with unity within the uncertainties. 
For ATLAS, NLO-to-data ratio was also shown in Refs.~\cite{Aad:2019,Aad:2023} lying below the NNLO-to-data ratio which might indicate the need for higher pQCD orders when the isolation momentum criterion is not tight enough to reproduce the contribution from fragmentation photons better. 

\begin{figure}[ht]
    \begin{center}
    \includegraphics[width=0.7\textwidth]{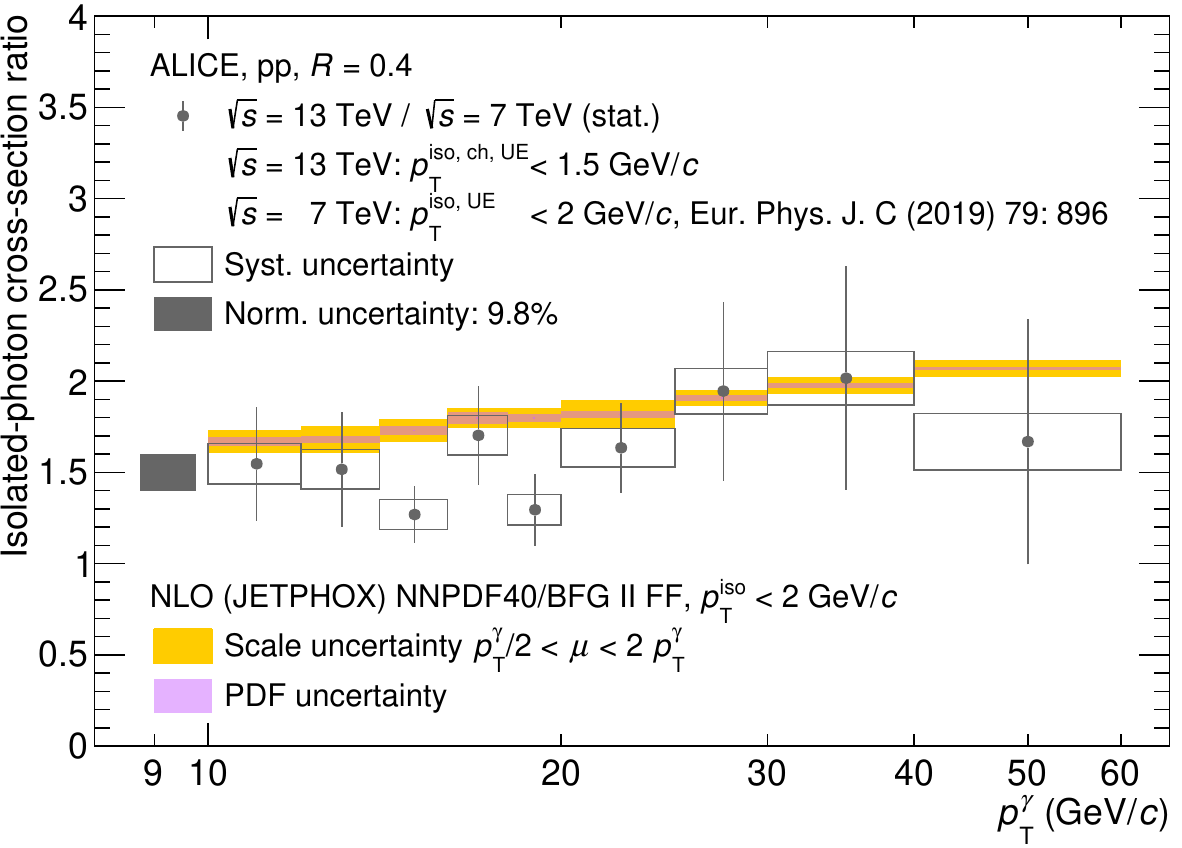}
    \end{center}
    \caption{\label{fig:13Over7} 
     (colour online) Ratio of isolated-photon cross sections measured in pp collisions at \s~=~13~TeV over the previous ALICE measurement at \s~=~7~TeV (taken from Ref.~\cite{ALICE:2019rtd}) compared to NLO calculations from JETPHOX shown as coloured boxes.
    }
\end{figure}

Figure~\ref{fig:13Over7} shows the ratio of the isolated-photon cross section measured by ALICE in pp collisions at \s~=~13~TeV to \s~=~7~TeV~\cite{ALICE:2019rtd} in data and NLO calculations. The previous ALICE measurement at \s~=~7~TeV is divided by the corresponding $\kappa^{\rm iso}$ since it was not done in the publication (the NLO calculation was multiplied instead by this factor). 
Almost all of the systematic uncertainties cancel in the ratio.
The contributions that are considered are: 
Half of the \s~=~13~TeV isolation probability; The full cluster--track matching uncertainty from \s~=~7~TeV measurement that accounts for the effect in the isolation-momentum calculation since clusters are used; The SM-dependent variation which does not cancel since it can vary in different data-taking periods. 
Considering a similar dispersion for the \s~=~7~TeV and 13~TeV samples, we take as uncertainty 4\%$\times \sqrt{2}$ = 5.6\%.
All these systematic sources are added in quadrature, resulting in an uncertainty on the ratio between 6.4\% and 9\% depending on \ptg. In any case, the dominant uncertainty is the statistical one that ranges from 15\% to 40\%. 
The NLO scale uncertainty decreases with \ptg from 4\% to 2\% and the PDF uncertainty varies between 1.4\% and 0.5\%.
The normalisation uncertainty is calculated from the luminosity of the two measurements and results in 9.8\%, dominated by the uncertainty on the integrated luminosity of the sample at \s~=~7~TeV.
In data, the ratio seems rather constant as a function of \ptg, showing a value of about 1.5, but is also compatible within the uncertainties to the NLO slow rise from close to 1.6 at 10--12~GeV/$c$ to close to 2.1 at 40--60~GeV/$c$. 
The qualitative agreement of the cross section ratio in data and pQCD calculation indicates that the underlying mechanisms in the theoretical approach are valid.

\begin{figure}[hb]
    \begin{center}
    \includegraphics[width=0.78\textwidth]{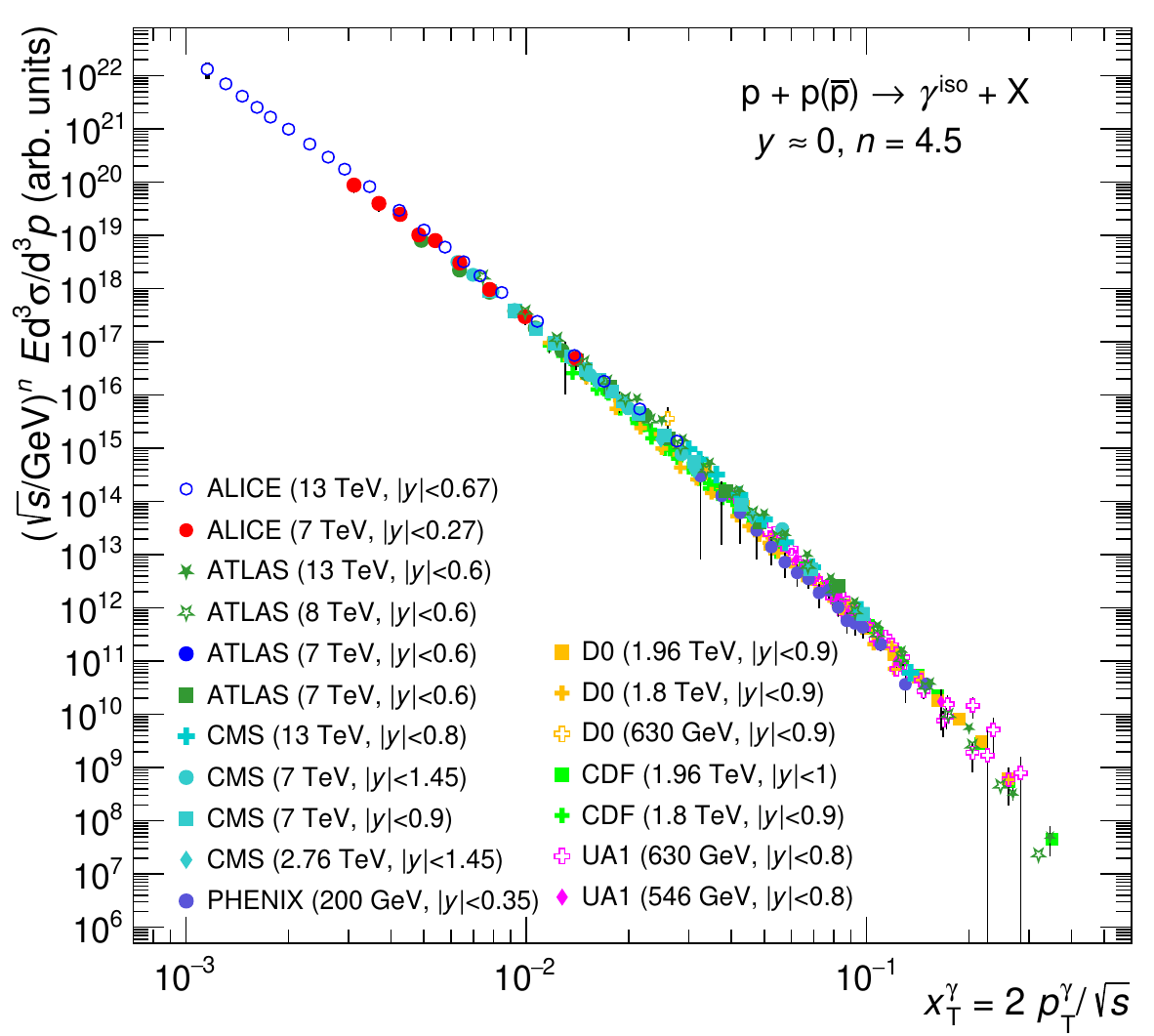}
    \end{center}
    \caption{\label{fig:isoPhotonWorld}(colour online) ALICE data compared to 
    previous measurements of isolated-photon spectra 
    in pp and p$\overline{\rm p}$ collisions as a function of $x_{\rm T }$ where the invariant cross sections have been scaled by $(\sqrt{s})^n$ with $n=4.5$ 
    (taken from Ref.~\cite{DENTERRIA2012311}). For this comparison, only the results covering midrapidity are shown.}
\end{figure}

For a comparison of cross sections measured at different \s, it is more appropriate to use the variable \xtg, defined earlier, which is also closely related to Bjorken $x$~\cite{Bjorken}.
A compilation of all available data on isolated-photon cross section measurements in collider experiments has been performed in~\cite{DENTERRIA2012311} and all \xtg\ spectra were compatible with a single curve when scaled by $(\sqrt{s})^n$ with $n=4.5$.
The ALICE measurements are compared to other measurements made at midrapidity including also those from other LHC experiments and the result is presented in Fig.~\ref{fig:isoPhotonWorld}. 
The ALICE measurement reported here, as anticipated, allows us to extend the \xtg\ reach to the lowest values measured at midrapidity so far, and is in agreement with the $n=4.5$ scaling, suggesting that all data are sensitive to the same production mechanisms. This measurement will help to constrain further the gluon PDF at midrapidity in the region $x\approx 1-3 \times 10^{-3}$ and reduce its uncertainty values.
However, the value $n=4.5$ deviates from the $1/(\ptg)^{n=4}$ dependence expected for the leading-twist partonic production mechanisms. 
This is due to effects like the running coupling and the evolution of PDFs, and indicates significant contributions from fragmentation photons and higher-twist diagrams~\cite{PhysRevLett.105.062002}. 

\section{Conclusions\label{sec:conclusion}}

The isolated-photon differential cross section in pp collisions at $\sqrt s = 13$~TeV was measured by the ALICE Collaboration at midrapidity in the transverse momentum range from 7 to 200~\GeVc.
Results are compared to ATLAS and CMS measurements and pQCD calculations. The mutual agreement of the data sets with theory supports the theoretical calculations and demonstrates the consistency of the different measurements.  

The current measurement extends the lower limit of {\ptg} compared to previous measurements by other LHC experiments, by an order of magnitude compared to ATLAS and CMS at the same collision energy. 
The measurement provides the lowest Bjorken-$x$ probed with isolated photons at midrapidity to date, showing an agreement between all the measurements with a common scale using $n=4.5$ over several orders of \xtg.  The low-$x$ measured will provide constraints on the gluon PDF. 

The lower \ptg\ reach of ALICE will be useful for future studies of isolated-photon cross sections and correlations of isolated photons to jets or hadrons used to constrain pQCD calculations, PDFs, and FFs, in particular, also for studying medium-induced modifications of hard probes in nucleus--nucleus collisions.

\clearpage


\newenvironment{acknowledgement}{\relax}{\relax}
\begin{acknowledgement}
\section*{Acknowledgements}

The authors would like to thank D. d’Enterria for fruitful discussions and the CMS Collaboration for providing details on previous measurements.


The ALICE Collaboration would like to thank all its engineers and technicians for their invaluable contributions to the construction of the experiment and the CERN accelerator teams for the outstanding performance of the LHC complex.
The ALICE Collaboration gratefully acknowledges the resources and support provided by all Grid centres and the Worldwide LHC Computing Grid (WLCG) collaboration.
The ALICE Collaboration acknowledges the following funding agencies for their support in building and running the ALICE detector:
A. I. Alikhanyan National Science Laboratory (Yerevan Physics Institute) Foundation (ANSL), State Committee of Science and World Federation of Scientists (WFS), Armenia;
Austrian Academy of Sciences, Austrian Science Fund (FWF): [M 2467-N36] and Nationalstiftung f\"{u}r Forschung, Technologie und Entwicklung, Austria;
Ministry of Communications and High Technologies, National Nuclear Research Center, Azerbaijan;
Conselho Nacional de Desenvolvimento Cient\'{\i}fico e Tecnol\'{o}gico (CNPq), Financiadora de Estudos e Projetos (Finep), Funda\c{c}\~{a}o de Amparo \`{a} Pesquisa do Estado de S\~{a}o Paulo (FAPESP) and Universidade Federal do Rio Grande do Sul (UFRGS), Brazil;
Bulgarian Ministry of Education and Science, within the National Roadmap for Research Infrastructures 2020-2027 (object CERN), Bulgaria;
Ministry of Education of China (MOEC) , Ministry of Science \& Technology of China (MSTC) and National Natural Science Foundation of China (NSFC), China;
Ministry of Science and Education and Croatian Science Foundation, Croatia;
Centro de Aplicaciones Tecnol\'{o}gicas y Desarrollo Nuclear (CEADEN), Cubaenerg\'{\i}a, Cuba;
Ministry of Education, Youth and Sports of the Czech Republic, Czech Republic;
The Danish Council for Independent Research | Natural Sciences, the VILLUM FONDEN and Danish National Research Foundation (DNRF), Denmark;
Helsinki Institute of Physics (HIP), Finland;
Commissariat \`{a} l'Energie Atomique (CEA) and Institut National de Physique Nucl\'{e}aire et de Physique des Particules (IN2P3) and Centre National de la Recherche Scientifique (CNRS), France;
Bundesministerium f\"{u}r Bildung und Forschung (BMBF) and GSI Helmholtzzentrum f\"{u}r Schwerionenforschung GmbH, Germany;
General Secretariat for Research and Technology, Ministry of Education, Research and Religions, Greece;
National Research, Development and Innovation Office, Hungary;
Department of Atomic Energy Government of India (DAE), Department of Science and Technology, Government of India (DST), University Grants Commission, Government of India (UGC) and Council of Scientific and Industrial Research (CSIR), India;
National Research and Innovation Agency - BRIN, Indonesia;
Istituto Nazionale di Fisica Nucleare (INFN), Italy;
Japanese Ministry of Education, Culture, Sports, Science and Technology (MEXT) and Japan Society for the Promotion of Science (JSPS) KAKENHI, Japan;
Consejo Nacional de Ciencia (CONACYT) y Tecnolog\'{i}a, through Fondo de Cooperaci\'{o}n Internacional en Ciencia y Tecnolog\'{i}a (FONCICYT) and Direcci\'{o}n General de Asuntos del Personal Academico (DGAPA), Mexico;
Nederlandse Organisatie voor Wetenschappelijk Onderzoek (NWO), Netherlands;
The Research Council of Norway, Norway;
Pontificia Universidad Cat\'{o}lica del Per\'{u}, Peru;
Ministry of Science and Higher Education, National Science Centre and WUT ID-UB, Poland;
Korea Institute of Science and Technology Information and National Research Foundation of Korea (NRF), Republic of Korea;
Ministry of Education and Scientific Research, Institute of Atomic Physics, Ministry of Research and Innovation and Institute of Atomic Physics and Universitatea Nationala de Stiinta si Tehnologie Politehnica Bucuresti, Romania;
Ministry of Education, Science, Research and Sport of the Slovak Republic, Slovakia;
National Research Foundation of South Africa, South Africa;
Swedish Research Council (VR) and Knut \& Alice Wallenberg Foundation (KAW), Sweden;
European Organization for Nuclear Research, Switzerland;
Suranaree University of Technology (SUT), National Science and Technology Development Agency (NSTDA) and National Science, Research and Innovation Fund (NSRF via PMU-B B05F650021), Thailand;
Turkish Energy, Nuclear and Mineral Research Agency (TENMAK), Turkey;
National Academy of  Sciences of Ukraine, Ukraine;
Science and Technology Facilities Council (STFC), United Kingdom;
National Science Foundation of the United States of America (NSF) and United States Department of Energy, Office of Nuclear Physics (DOE NP), United States of America.
In addition, individual groups or members have received support from:
Czech Science Foundation (grant no. 23-07499S), Czech Republic;
European Research Council (grant no. 950692), European Union;
ICSC - Centro Nazionale di Ricerca in High Performance Computing, Big Data and Quantum Computing, European Union - NextGenerationEU;
Academy of Finland (Center of Excellence in Quark Matter) (grant nos. 346327, 346328), Finland.

\end{acknowledgement}

\bibliographystyle{utphys}   
\bibliography{bibliography}

\newpage
\appendix

\section{The ALICE Collaboration}
\label{app:collab}
\begin{flushleft} 
\small

S.~Acharya\,\orcidlink{0000-0002-9213-5329}\,$^{\rm 127}$, 
D.~Adamov\'{a}\,\orcidlink{0000-0002-0504-7428}\,$^{\rm 86}$, 
A.~Agarwal$^{\rm 135}$, 
G.~Aglieri Rinella\,\orcidlink{0000-0002-9611-3696}\,$^{\rm 32}$, 
L.~Aglietta\,\orcidlink{0009-0003-0763-6802}\,$^{\rm 24}$, 
M.~Agnello\,\orcidlink{0000-0002-0760-5075}\,$^{\rm 29}$, 
N.~Agrawal\,\orcidlink{0000-0003-0348-9836}\,$^{\rm 25}$, 
Z.~Ahammed\,\orcidlink{0000-0001-5241-7412}\,$^{\rm 135}$, 
S.~Ahmad\,\orcidlink{0000-0003-0497-5705}\,$^{\rm 15}$, 
S.U.~Ahn\,\orcidlink{0000-0001-8847-489X}\,$^{\rm 71}$
% ...

\section*{Affiliation Notes}

$^{\rm I}$ Deceased\\
$^{\rm II}$ Also at: Max-Planck-Institut fur Physik, Munich, Germany\\
$^{\rm III}$ Also at: Italian National Agency for New Technologies, Energy and Sustainable Economic Development (ENEA), Bologna, Italy\\
$^{\rm IV}$ Also at: Dipartimento DET del Politecnico di Torino, Turin, Italy\\
$^{\rm V}$ Also at: Yildiz Technical University, Istanbul, T\"{u}rkiye\\
$^{\rm VI}$ Also at: Department of Applied Physics, Aligarh Muslim University, Aligarh, India\\
$^{\rm VII}$ Also at: Institute of Theoretical Physics, University of Wroclaw, Poland\\
$^{\rm VIII}$ Also at: An institution covered by a cooperation agreement with CERN\\

\section*{Collaboration Institutes}

$^{1}$ A.I. Alikhanyan National Science Laboratory (Yerevan Physics Institute) Foundation, Yerevan, Armenia\\
$^{2}$ AGH University of Krakow, Cracow, Poland\\
$^{3}$ Bogolyubov Institute for Theoretical Physics, National Academy of Sciences of Ukraine, Kiev, Ukraine\\
$^{4}$ Bose Institute, Department of Physics  and Centre for Astroparticle Physics and Space Science (CAPSS), Kolkata, India\\
$^{5}$ California Polytechnic State University, San Luis Obispo, California, United States\\
$^{6}$ Central China Normal University, Wuhan, China\\
$^{7}$ Centro de Aplicaciones Tecnol\'{o}gicas y Desarrollo Nuclear (CEADEN), Havana, Cuba\\
$^{8}$ Centro de Investigaci\'{o}n y de Estudios Avanzados (CINVESTAV), Mexico City and M\'{e}rida, Mexico\\
$^{9}$ Chicago State University, Chicago, Illinois, United States\\
$^{10}$ China Institute of Atomic Energy, Beijing, China\\
$^{11}$ China University of Geosciences, Wuhan, China\\
$^{12}$ Chungbuk National University, Cheongju, Republic of Korea\\
$^{13}$ Comenius University Bratislava, Faculty of Mathematics, Physics and Informatics, Bratislava, Slovak Republic\\
$^{14}$ Creighton University, Omaha, Nebraska, United States\\
$^{15}$ Department of Physics, Aligarh Muslim University, Aligarh, India\\
$^{16}$ Department of Physics, Pusan National University, Pusan, Republic of Korea\\
$^{17}$ Department of Physics, Sejong University, Seoul, Republic of Korea\\
$^{18}$ Department of Physics, University of California, Berkeley, California, United States\\
$^{19}$ Department of Physics, University of Oslo, Oslo, Norway\\
$^{20}$ Department of Physics and Technology, University of Bergen, Bergen, Norway\\
$^{21}$ Dipartimento di Fisica, Universit\`{a} di Pavia, Pavia, Italy\\
$^{22}$ Dipartimento di Fisica dell'Universit\`{a} and Sezione INFN, Cagliari, Italy\\
$^{23}$ Dipartimento di Fisica dell'Universit\`{a} and Sezione INFN, Trieste, Italy\\
$^{24}$ Dipartimento di Fisica dell'Universit\`{a} and Sezione INFN, Turin, Italy\\
$^{25}$ Dipartimento di Fisica e Astronomia dell'Universit\`{a} and Sezione INFN, Bologna, Italy\\
$^{26}$ Dipartimento di Fisica e Astronomia dell'Universit\`{a} and Sezione INFN, Catania, Italy\\
$^{27}$ Dipartimento di Fisica e Astronomia dell'Universit\`{a} and Sezione INFN, Padova, Italy\\
$^{28}$ Dipartimento di Fisica `E.R.~Caianiello' dell'Universit\`{a} and Gruppo Collegato INFN, Salerno, Italy\\
$^{29}$ Dipartimento DISAT del Politecnico and Sezione INFN, Turin, Italy\\
$^{30}$ Dipartimento di Scienze MIFT, Universit\`{a} di Messina, Messina, Italy\\
$^{31}$ Dipartimento Interateneo di Fisica `M.~Merlin' and Sezione INFN, Bari, Italy\\
$^{32}$ European Organization for Nuclear Research (CERN), Geneva, Switzerland\\
$^{33}$ Faculty of Electrical Engineering, Mechanical Engineering and Naval Architecture, University of Split, Split, Croatia\\
$^{34}$ Faculty of Engineering and Science, Western Norway University of Applied Sciences, Bergen, Norway\\
$^{35}$ Faculty of Nuclear Sciences and Physical Engineering, Czech Technical University in Prague, Prague, Czech Republic\\
$^{36}$ Faculty of Physics, Sofia University, Sofia, Bulgaria\\
$^{37}$ Faculty of Science, P.J.~\v{S}af\'{a}rik University, Ko\v{s}ice, Slovak Republic\\
$^{38}$ Frankfurt Institute for Advanced Studies, Johann Wolfgang Goethe-Universit\"{a}t Frankfurt, Frankfurt, Germany\\
$^{39}$ Fudan University, Shanghai, China\\
$^{40}$ Gangneung-Wonju National University, Gangneung, Republic of Korea\\
$^{41}$ Gauhati University, Department of Physics, Guwahati, India\\
$^{42}$ Helmholtz-Institut f\"{u}r Strahlen- und Kernphysik, Rheinische Friedrich-Wilhelms-Universit\"{a}t Bonn, Bonn, Germany\\
$^{43}$ Helsinki Institute of Physics (HIP), Helsinki, Finland\\
$^{44}$ High Energy Physics Group,  Universidad Aut\'{o}noma de Puebla, Puebla, Mexico\\
$^{45}$ Horia Hulubei National Institute of Physics and Nuclear Engineering, Bucharest, Romania\\
$^{46}$ HUN-REN Wigner Research Centre for Physics, Budapest, Hungary\\
$^{47}$ Indian Institute of Technology Bombay (IIT), Mumbai, India\\
$^{48}$ Indian Institute of Technology Indore, Indore, India\\
$^{49}$ INFN, Laboratori Nazionali di Frascati, Frascati, Italy\\
$^{50}$ INFN, Sezione di Bari, Bari, Italy\\
$^{51}$ INFN, Sezione di Bologna, Bologna, Italy\\
$^{52}$ INFN, Sezione di Cagliari, Cagliari, Italy\\
$^{53}$ INFN, Sezione di Catania, Catania, Italy\\
$^{54}$ INFN, Sezione di Padova, Padova, Italy\\
$^{55}$ INFN, Sezione di Pavia, Pavia, Italy\\
$^{56}$ INFN, Sezione di Torino, Turin, Italy\\
$^{57}$ INFN, Sezione di Trieste, Trieste, Italy\\
$^{58}$ Inha University, Incheon, Republic of Korea\\
$^{59}$ Institute for Gravitational and Subatomic Physics (GRASP), Utrecht University/Nikhef, Utrecht, Netherlands\\
$^{60}$ Institute of Experimental Physics, Slovak Academy of Sciences, Ko\v{s}ice, Slovak Republic\\
$^{61}$ Institute of Physics, Homi Bhabha National Institute, Bhubaneswar, India\\
$^{62}$ Institute of Physics of the Czech Academy of Sciences, Prague, Czech Republic\\
$^{63}$ Institute of Space Science (ISS), Bucharest, Romania\\
$^{64}$ Institut f\"{u}r Kernphysik, Johann Wolfgang Goethe-Universit\"{a}t Frankfurt, Frankfurt, Germany\\
$^{65}$ Instituto de Ciencias Nucleares, Universidad Nacional Aut\'{o}noma de M\'{e}xico, Mexico City, Mexico\\
$^{66}$ Instituto de F\'{i}sica, Universidade Federal do Rio Grande do Sul (UFRGS), Porto Alegre, Brazil\\
$^{67}$ Instituto de F\'{\i}sica, Universidad Nacional Aut\'{o}noma de M\'{e}xico, Mexico City, Mexico\\
$^{68}$ iThemba LABS, National Research Foundation, Somerset West, South Africa\\
$^{69}$ Jeonbuk National University, Jeonju, Republic of Korea\\
$^{70}$ Johann-Wolfgang-Goethe Universit\"{a}t Frankfurt Institut f\"{u}r Informatik, Fachbereich Informatik und Mathematik, Frankfurt, Germany\\
$^{71}$ Korea Institute of Science and Technology Information, Daejeon, Republic of Korea\\
$^{72}$ KTO Karatay University, Konya, Turkey\\
$^{73}$ Laboratoire de Physique Subatomique et de Cosmologie, Universit\'{e} Grenoble-Alpes, CNRS-IN2P3, Grenoble, France\\
$^{74}$ Lawrence Berkeley National Laboratory, Berkeley, California, United States\\
$^{75}$ Lund University Department of Physics, Division of Particle Physics, Lund, Sweden\\
$^{76}$ Nagasaki Institute of Applied Science, Nagasaki, Japan\\
$^{77}$ Nara Women{'}s University (NWU), Nara, Japan\\
$^{78}$ National and Kapodistrian University of Athens, School of Science, Department of Physics , Athens, Greece\\
$^{79}$ National Centre for Nuclear Research, Warsaw, Poland\\
$^{80}$ National Institute of Science Education and Research, Homi Bhabha National Institute, Jatni, India\\
$^{81}$ National Nuclear Research Center, Baku, Azerbaijan\\
$^{82}$ National Research and Innovation Agency - BRIN, Jakarta, Indonesia\\
$^{83}$ Niels Bohr Institute, University of Copenhagen, Copenhagen, Denmark\\
$^{84}$ Nikhef, National institute for subatomic physics, Amsterdam, Netherlands\\
$^{85}$ Nuclear Physics Group, STFC Daresbury Laboratory, Daresbury, United Kingdom\\
$^{86}$ Nuclear Physics Institute of the Czech Academy of Sciences, Husinec-\v{R}e\v{z}, Czech Republic\\
$^{87}$ Oak Ridge National Laboratory, Oak Ridge, Tennessee, United States\\
$^{88}$ Ohio State University, Columbus, Ohio, United States\\
$^{89}$ Physics department, Faculty of science, University of Zagreb, Zagreb, Croatia\\
$^{90}$ Physics Department, Panjab University, Chandigarh, India\\
$^{91}$ Physics Department, University of Jammu, Jammu, India\\
$^{92}$ Physics Program and International Institute for Sustainability with Knotted Chiral Meta Matter (SKCM2), Hiroshima University, Hiroshima, Japan\\
$^{93}$ Physikalisches Institut, Eberhard-Karls-Universit\"{a}t T\"{u}bingen, T\"{u}bingen, Germany\\
$^{94}$ Physikalisches Institut, Ruprecht-Karls-Universit\"{a}t Heidelberg, Heidelberg, Germany\\
$^{95}$ Physik Department, Technische Universit\"{a}t M\"{u}nchen, Munich, Germany\\
$^{96}$ Politecnico di Bari and Sezione INFN, Bari, Italy\\
$^{97}$ Research Division and ExtreMe Matter Institute EMMI, GSI Helmholtzzentrum f\"ur Schwerionenforschung GmbH, Darmstadt, Germany\\
$^{98}$ Saga University, Saga, Japan\\
$^{99}$ Saha Institute of Nuclear Physics, Homi Bhabha National Institute, Kolkata, India\\
$^{100}$ School of Physics and Astronomy, University of Birmingham, Birmingham, United Kingdom\\
$^{101}$ Secci\'{o}n F\'{\i}sica, Departamento de Ciencias, Pontificia Universidad Cat\'{o}lica del Per\'{u}, Lima, Peru\\
$^{102}$ Stefan Meyer Institut f\"{u}r Subatomare Physik (SMI), Vienna, Austria\\
$^{103}$ SUBATECH, IMT Atlantique, Nantes Universit\'{e}, CNRS-IN2P3, Nantes, France\\
$^{104}$ Sungkyunkwan University, Suwon City, Republic of Korea\\
$^{105}$ Suranaree University of Technology, Nakhon Ratchasima, Thailand\\
$^{106}$ Technical University of Ko\v{s}ice, Ko\v{s}ice, Slovak Republic\\
$^{107}$ The Henryk Niewodniczanski Institute of Nuclear Physics, Polish Academy of Sciences, Cracow, Poland\\
$^{108}$ The University of Texas at Austin, Austin, Texas, United States\\
$^{109}$ Universidad Aut\'{o}noma de Sinaloa, Culiac\'{a}n, Mexico\\
$^{110}$ Universidade de S\~{a}o Paulo (USP), S\~{a}o Paulo, Brazil\\
$^{111}$ Universidade Estadual de Campinas (UNICAMP), Campinas, Brazil\\
$^{112}$ Universidade Federal do ABC, Santo Andre, Brazil\\
$^{113}$ Universitatea Nationala de Stiinta si Tehnologie Politehnica Bucuresti, Bucharest, Romania\\
$^{114}$ University of Cape Town, Cape Town, South Africa\\
$^{115}$ University of Derby, Derby, United Kingdom\\
$^{116}$ University of Houston, Houston, Texas, United States\\
$^{117}$ University of Jyv\"{a}skyl\"{a}, Jyv\"{a}skyl\"{a}, Finland\\
$^{118}$ University of Kansas, Lawrence, Kansas, United States\\
$^{119}$ University of Liverpool, Liverpool, United Kingdom\\
$^{120}$ University of Science and Technology of China, Hefei, China\\
$^{121}$ University of South-Eastern Norway, Kongsberg, Norway\\
$^{122}$ University of Tennessee, Knoxville, Tennessee, United States\\
$^{123}$ University of the Witwatersrand, Johannesburg, South Africa\\
$^{124}$ University of Tokyo, Tokyo, Japan\\
$^{125}$ University of Tsukuba, Tsukuba, Japan\\
$^{126}$ Universit\"{a}t M\"{u}nster, Institut f\"{u}r Kernphysik, M\"{u}nster, Germany\\
$^{127}$ Universit\'{e} Clermont Auvergne, CNRS/IN2P3, LPC, Clermont-Ferrand, France\\
$^{128}$ Universit\'{e} de Lyon, CNRS/IN2P3, Institut de Physique des 2 Infinis de Lyon, Lyon, France\\
$^{129}$ Universit\'{e} de Strasbourg, CNRS, IPHC UMR 7178, F-67000 Strasbourg, France, Strasbourg, France\\
$^{130}$ Universit\'{e} Paris-Saclay, Centre d'Etudes de Saclay (CEA), IRFU, D\'{e}partment de Physique Nucl\'{e}aire (DPhN), Saclay, France\\
$^{131}$ Universit\'{e}  Paris-Saclay, CNRS/IN2P3, IJCLab, Orsay, France\\
$^{132}$ Universit\`{a} degli Studi di Foggia, Foggia, Italy\\
$^{133}$ Universit\`{a} del Piemonte Orientale, Vercelli, Italy\\
$^{134}$ Universit\`{a} di Brescia, Brescia, Italy\\
$^{135}$ Variable Energy Cyclotron Centre, Homi Bhabha National Institute, Kolkata, India\\
$^{136}$ Warsaw University of Technology, Warsaw, Poland\\
$^{137}$ Wayne State University, Detroit, Michigan, United States\\
$^{138}$ Yale University, New Haven, Connecticut, United States\\
$^{139}$ Yonsei University, Seoul, Republic of Korea\\
$^{140}$  Zentrum  f\"{u}r Technologie und Transfer (ZTT), Worms, Germany\\
$^{141}$ Affiliated with an institute covered by a cooperation agreement with CERN\\
$^{142}$ Affiliated with an international laboratory covered by a cooperation agreement with CERN.\\

\end{flushleft} 

\end{document}